\documentclass[11pt]{article}
\usepackage[utf8]{inputenc}
\usepackage{graphicx}
\usepackage{amsmath}
\usepackage{amssymb}
\usepackage{cancel}
\usepackage{empheq} 
\usepackage{braket}
\usepackage{slashed}
\usepackage{float}
\usepackage[footnotesize]{caption}
\usepackage{subcaption}
\usepackage{scalerel}
\usepackage{empheq}
\graphicspath{ {images/} }
\usepackage[a4paper,width=150mm,top=25mm,bottom=25mm]{geometry}
\usepackage{fancyhdr}
\usepackage{upgreek}
\usepackage[dvipsnames]{xcolor}

\usepackage{jheppub}
\newcommand{\mycomment}[1]{}

\title{From BTZ Perturbations to Schwarzian Modes: A Geometrical and Perturbative Analysis}
\author[a]{Lucas Acito}

\author[b]{and Matías N. Sempé}

\affiliation[a]{{\it Instituto de F\'\i sica La Plata (IFLP), CONICET  \& 
Departmento de F\'\i sica  Dr. Emil H. Bose, UNLP, 
C.C. 67, (1900) La Plata, Argentina.}}

\affiliation[b]{Instituto de Astronomía y Física del Espacio, (CONICET-UBA) Ciudad Universitaria, Pabellón IAFE, CABA, C1428ZAA, Argentina}

\emailAdd{acitolucas@iflp.unlp.edu.ar}
\emailAdd{sempe\_100@hotmail.com}

\abstract{
We provide a detailed derivation of the Schwarzian modes in the full geometry of the Bañados-Teitelboim-Zanelli (BTZ) black hole at finite temperature, establishing the precise conditions under which they emerge from the general solution, thereby clarifying the absence of rotational modes in the full geometry. In addition, we demonstrate that the same modes can be recovered through a purely geometric Kerr–Schild construction. This equivalent approach offers a new geometric understanding of the Schwarzian sector and highlights the correspondence between perturbative and pure geometric approaches, additionally it provides a connection with double copy.
}

\begin{document}
\maketitle
\section{Introduction}

The thermodynamic nature of black holes reveals a remarkable interplay between gravity and quantum theory, with the Euclidean path integral approach providing a concrete framework for its study. Classical saddle point computations first showed that the entropy is captured, to leading order, by the horizon area in Planck units, with the temperature determined by the surface gravity \cite{Hawking:1977}. Historically, these semiclassical computations, together with the earliest perturbative analyses of black hole geometries at low temperatures, suggested the presence of a large degeneracy \cite{Michelson:1999kn, Page:2000dk}. However, exact computations in supergravity models show, through microstate counting, that the ground states is not degenerate and that there is a logarithmic correction to the entropy \cite{Gupta:2011(1), Gupta:2011(2), Sen:2012kpz}. These results for extremal black hole degeneracy, nevertheless, rely crucially on supersymmetry and may not directly apply once it is broken.

In recent years, rotating black holes at low temperatures have played a central role in the discussions of microscopic origin of non-supersymmetric black hole entropy \cite{Sen:2012cj, Sen:2012dw}. Perturbative analyses around the near-extremal geometry reveal quantum effects, including logarithmic corrections to the entropy, indicating that these black holes do not exhibit a macroscopic degeneracy, but instead behave as ordinary quantum systems with a discrete spectrum \cite{Ghosh:2019rcj, Iliesiu:2020qvm, Iliesiu:2022onk}.

The degeneracy problem was solved by the existence of the Schwarzians, modes that are eigenfunctions of the Lichnerowicz operator with a temperature dependant eigenvalue which goes to zero at the extremal limit \cite{Sen:2012dw}. In particular, these studies have shown that the low-temperature dynamics is governed by the Schwarzian corrections, which arise from the breaking of the near-horizon symmetry group 
$SL(2,R)$ \cite{Kunduri:2007vf}. The formalism for computing these corrections was systematically developed in the work by Sen and collaborators \cite{Sen:2012cj,Sen:2012dw}, providing a powerful framework to study the statistical mechanics of near-extremal black holes.

The discovery of the Schwarzian modes led to an important conceptual shift: non-supersymmetric black holes, which at first seemed similar to their supersymmetric counterparts at low temperatures, actually display a very different structure once these corrections are properly taken into account \cite{Kapec:2023ruw,Cassani:2025sim,Leo,Joacokerr}. Nevertheless, most of the existing analyses rely heavily on the near-horizon extremal geometry and treat finite-temperature effects only perturbatively. This leaves open the question of whether Schwarzian modes can be consistently identified in the full black hole geometry beyond the near-horizon regime, a clarification that would also shed light on how different near-extremal limits produce distinct gravitational features \cite{tanos1}.

In this work, we aim to advance this direction by extending the analysis of \cite{JOACO,Castro:2025itb}, who initiated the study of Schwarzian modes in the complete geometry. Our focus will be on the BTZ black hole \cite{Banados:1992wn}, which provides an ideal testing ground: not only does it admit a fully analytic treatment \cite{Castro:2017mfj,Indios,Leston}, but it also encapsulates many of the essential features of higher-dimensional rotating black holes. By studying this case, we are able to clearly identify the conditions under which Schwarzian modes appear and how they are embedded in the full spectrum of fluctuations.

Our main results are threefold. First, we demonstrate that the graviton sector admits a finite number of normalizable Schwarzian modes that exhibit an asymptotic behavior more general than the standard Brown-Henneaux boundary conditions \cite{Grumiller}. Second, we show that the normalizable Schwarzian modes obtained in the full geometry smoothly reduce to the well-known near-horizon modes, becoming an infinite tower of modes in the extremal limit, thus providing a consistent bridge between the two perspectives. Third, we show that the same Schwarzian modes can also be obtained from a purely geometrical construction using a Kerr–Schild metric, where the vector in this decomposition corresponds to the Schwarzian sector of the vector field. This introduces a general procedure for identifying Schwarzian modes in broader classes of backgrounds and establishes a direct link between these modes and gauge field theories.

The correspondence Gravity=(Yang-Mills)$^2$ was first observed in the context of scattering amplitudes, where gravitational amplitudes can be constructed from the corresponding gluon amplitudes \cite{Del_Duca:2000}. Its first explicit realization at the level of perturbative gravity and gauge theories, known as the double copy, was developed in \cite{Monteiro:2014}. In three-dimensional gravity, the Kerr-Schild metric can be used around $AdS_3$ to interpret the BTZ black hole as a gauge theory with a source \cite{Carrillo:2018,Carrillo:2019}. Therefore, if the Schwarzian modes can be derived from a Kerr-Schild construction around BTZ, it is natural to expect a similar interplay between these modes and the double copy framework. 

The structure of the paper is as follows. In Section \ref{sec:btz}, we review the main features of the BTZ geometry relevant to our analysis. Section \ref{sec:spin2} is devoted to the study of spin-2 perturbations, where we explicitly compute the fluctuation spectrum and identify all the normalizable modes (the Schwarzian, the rotational, and other non-zero modes). In Section \ref{sec:modes}, we analyze the obtained modes, establish their connection with the near-horizon dynamics, and compute their contribution to the one-loop path integral. In Section \ref{sec:Kerr_Schild}, we adopt a geometric perspective and construct the Kerr–Schild metric around the BTZ background, showing that the associated null and geodesic vectors coincide with the perturbatively obtained Schwarzian modes, which in turn coincide with the Schwarzian modes of the vector field computed in Appendix~\ref{sec:vector}. Finally, we conclude with a discussion of the implications of our results and possible extensions.

\section{BTZ geometry}\label{sec:btz}

In this work we will be concerned about the Euclidean theory of three dimensional gravity with a negative cosmological constant coupled to matter, with an action 
\begin{equation}
    I=-\frac{1}{16 \pi G_N} \int_M d^3x \:\sqrt{g} \left(R+\frac{2}{\ell^2}\right)-\frac{1}{8 \pi G_N} \oint_{\partial M} d^2x \:\sqrt{\gamma}\: K + I_{ct} + I_\textbf{matter}
\end{equation}
where the second term is the Gibbons-Hawking-York term defined by the intrinsic metric $\gamma_{\mu\nu}$ at the boundary, and $I_{ct}$ are the counter-terms for asymptotically $AdS_3$ spaces \cite{Balasubramanian:1999}. In the absence of matter, one of the most famous solution of this background is the rotating BTZ black hole \cite{BTZ}, which, in the Euclidean case and setting $\ell=1$ for simplicity, is given by  
\begin{equation}\label{BTZ_metric}
    ds^2=f(r) dt^2+\frac{dr^2}{f(r)}+r^2\left(d\phi+i \frac{r_+ r_-}{r^2}dt\right)^2
\end{equation}
with the lapse function $f(r)=(r^2-r^2_+)(r^2-r^2_-)/r^2$, where $r_-$ and $r_+$ are the inner and outer horizon respectively (both are reals). This solution is locally equivalent to $AdS_3$, but is topologically different as impose the following identifications $\phi \sim \phi+2\pi$ and $t \sim t+ T^{-1}$, where the temperature $T$ is given by
\begin{equation}
    T=\frac{r^2_+-r^2_-}{2 \pi r_+}
\end{equation}
Alternatively, one may define the left $T_L$ and right $T_R$ moving temperatures as
\begin{equation}\label{eq:T_LR}
    T_L=\frac{r_+-r_-}{2\pi}\:,\qquad T_R=\frac{r_++r_-}{2\pi}
\end{equation}
Another useful line element is
\begin{equation}
    ds^2=\frac{1}{4z(1-z)^2}dz^2+ \frac{z}{1-z}dx^2_-+\frac{1}{1-z} dx^2_+
    \label{Leston}
\end{equation}
both are related by 
\begin{equation}\label{btz_transformations}
    z=\frac{r^2-r^2_+}{r^2-r^2_-} \qquad,\qquad x_-=r_+ t -i\: r_- \phi \qquad\&\qquad x_+=- i \:r_- t-r_+ \phi 
\end{equation}

We can also obtain the black hole mass and angular momentum
\begin{align}
    M=\frac{r^2_++r^2_-}{8 G_N}\:, \qquad  J=\frac{r_+ r_-}{4 G_N}
\end{align}
The entropy at low temperature is given by \cite{Kapec:2024zdj}
\begin{equation}
    S=\frac{\pi}{2} \sqrt{\frac{J}{2}}+\frac{\pi^2}{4} T+ O(T^2)
\end{equation}
Notice that the first term is not zero at $r_+=r_-$ showing a high degeneracy, this is the term we want to correct using the Schwarzians.



\section{Spin-2 Field Perturbation}\label{sec:spin2}

Our goal is to determine the modes of the quadratic operator \cite{Lichne1,Lichne2} governing the dynamics to spin-2 fluctuations, defined from the perturbation $\tilde{g}_{\mu \nu}=g_{\mu \nu}+h_{\mu \nu}$ with $g_{\mu\nu}$ the Euclidean BTZ metric \eqref{BTZ_metric}. In the decoupled case, and upon imposing the transverse $h_{\mu \nu} g^{\mu \nu}=0$ and traceless gauge $\nabla_\mu h^{\mu \nu}=0$ conditions, the Lichnerowicz operator we need to diagonalize reduces to
\begin{equation}\label{graviton_lich}
    \left(-\nabla^2-2 \right)h_{\mu\nu} =(1+k^2)\: h_{\mu\nu}
\end{equation}
This equation can be factorized in terms of a first-order differential equation \cite{Vasiliev:1997}\footnote{The existence of both first- and second-order equations is related to the fact that, locally, the geometry is Euclidean $AdS_3$, where these operators correspond to the Casimirs. The same structure appears on 
$S^3$
 \cite{RiosFukelman:2023mgq}.}, given by
\begin{equation}\label{graviton_linear}
\epsilon_\mu^{\:\:\alpha\beta}\nabla_\alpha h_{\beta\nu} = \pm k\: h_{\mu\nu}
\end{equation}
It is instructive to show explicitly that \eqref{graviton_linear} implies \eqref{graviton_lich}. Applying the linear operator twice yields
\begin{equation} \label{graviton_proof_op}
    \begin{aligned}
        \epsilon_\mu^{\ \alpha \beta} \nabla_\alpha \left(\epsilon_\beta^{\ \gamma \sigma} \nabla_\gamma h_{\sigma \nu}\right)=-3 h_{\mu \nu}-\nabla^\alpha \nabla_\alpha h_{\mu \nu}=k^2 h_{\mu \nu} 
    \end{aligned}
\end{equation}
which implies
\begin{equation}
    \nabla^\alpha \nabla_\alpha h_{\mu \nu}=-(k^2+3) h_{\mu \nu}
\end{equation}
Moreover, non-zero modes of\eqref{graviton_linear} must satisfy $\nabla^\mu h_{\mu \nu}=0$ and $h_\mu^{\:\mu}=0$. The gauge fixing conditions follows from\eqref{graviton_linear}
 \begin{equation}
    k \nabla^\mu h_{\mu\nu}=\epsilon^{\mu\alpha\beta} \nabla_\mu\nabla_\alpha h_{\beta\nu} = \frac12 \epsilon^{\mu\alpha\beta} (R_{\beta \ \mu \alpha}^{\ \lambda } h_{\lambda \nu}+R_{\nu \ \mu \alpha}^{\ \lambda } h_{\lambda \beta})=0
 \end{equation} 
where in the last step we used the fact that in three dimensions $R_{abcd}=-(g_{ac} g_{bd}-g_{ad} g_{bc})$.\\
Instead of solving only\eqref{graviton_lich}, it is convenient to also use\eqref{graviton_linear}. This allows us to obtain relations among the tensor components
\begin{equation}\label{graviton_zcomponents}
    \begin{aligned}
        h_{zz}&= \frac{h_{--} +z\: h_{++}}{4 (z-1) z^2}
        \\ 
        h_{z+}&= \frac{\partial_+ h_{--} -\partial_-h_{-+} \mp k\:\partial_+h_{-+} \pm k\:\partial_- h_{++}}{2\left(k^2+1\right) z}
        \\
        h_{z-}&= \frac{\partial_- h_{++} -\partial_{+}h_{-+} \pm k\:\partial_{-}h_{-+} \mp k\:\partial_+ h_{--}}{2\left(k^2+1\right) z} 
    \end{aligned}
\end{equation}
Thus, only three independent components remain, namely $h_{--}$, $h_{-+}$ and $h_{++}$, from which all others can be reconstructed. Expanding the Laplacian on these independent components gives 
\begin{equation}
    \begin{pmatrix}
        \Box h_{++}\\
        \Box h_{-+}\\
        \Box h_{--}
    \end{pmatrix} = \begin{pmatrix}
        \Delta h_{++}\\
        \Delta h_{-+}\\
        \Delta h_{--}
    \end{pmatrix} + \begin{pmatrix}
        -4 & \pm4\:k & 2 \\
        \mp2\:k & -6 & \pm2\:k \\
        2 & \mp4\:k & -4
    \end{pmatrix} \begin{pmatrix}
        h_{++}\\
        h_{-+}\\
        h_{--}
    \end{pmatrix} 
\end{equation}
where the upper/lower signs correspond to the $\pm$ choice in the linear operator \eqref{graviton_linear}, and the scalar Laplacian is defined as $\Delta h_{\mu\nu} =\partial_\alpha \left(\sqrt{g}\:\partial^\alpha h_{\mu\nu}\right)/\sqrt{g}$.
Diagonalizing the coefficient matrix, we obtain new tensor components such that the Laplacian acts diagonally, effectively reducing the problem to three scalar equations.

Diagonalizing the coefficient matrix, we obtain new tensor components such that decouple this equations, reducing the problem to three scalar fields equations
\begin{equation}
    \begin{pmatrix}
        \Box h_{11}\\
        \Box h_{12}\\
        \Box h_{22}
    \end{pmatrix} = \begin{pmatrix}
        \Delta h_{11}\\
        \Delta h_{12}\\
        \Delta h_{22}
    \end{pmatrix} + \begin{pmatrix}
        -2 &     0      &  0 \\
         0 & -6-4i \:k  &  0 \\
         0 &     0      & -6+4i\:k
    \end{pmatrix} \begin{pmatrix}
        h_{11}\\
        h_{12}\\
        h_{22}
    \end{pmatrix} 
\end{equation}
where the transformation relating the two bases is
\begin{equation}\label{graviton_transformation}
    \begin{pmatrix}
        h_{++}\\
        h_{-+}\\
        h_{--}
    \end{pmatrix} = \begin{pmatrix}
        1 & -1 & -1 \\
        0 & -i &  i \\
        1 &  1 &  1
    \end{pmatrix} \begin{pmatrix}
        h_{11}\\
        h_{12}\\
        h_{22}
    \end{pmatrix}
\end{equation}
This matches the diagonalization procedure of \cite{Indios}, up to a Wick rotation. Hence the components $h_{++}$, $h_{-+}$ and $h_{--}$ are linear combinations of $h_{11}$, $h_{12}$ and $h_{22}$. It is worth stressing that the diagonalized quadratic operator does not depend on the sign choice in \eqref{graviton_linear}, and neither the transformation matrix \eqref{graviton_transformation}. This follows from \eqref{graviton_proof_op}, since the Laplacian is quadratic in the linear operator. However, the sign in \eqref{graviton_linear} will become relevant when we determine the polarization structure. 
\\
The resulting decoupled equations are
\begin{equation}\label{graviton_eqns}
    \left\{\begin{aligned}
        \Delta h_{11} + (1+k^2)h_{11} = 0  \\
        \Delta h_{12} + \left(k(k-4i) -3\right)h_{12} = 0 \\
        \Delta h_{22} + (i+k)(3i+k) h_{22} = 0
    \end{aligned}\right.
\end{equation}
Thus we are left with three scalar-like fields equations. To solve them we use the ansatz of \cite{Leston,Indios}
\begin{equation}\label{eq:anzat}
    h_{ij}=e^{i(\kappa_+ x^+ + \kappa_- x^-)} R_{ij}(z) \qquad i,\:j=1,\:2
\end{equation}
Additionally, we require these modes to be Matsubara modes when transformed back to BTZ coordinates via \eqref{btz_transformations}. We focus on modes that depend only on the Euclidean time in the Matsubara expansion, implying that $\kappa_-$ and $\kappa_+$ are not independent\footnote{Here we fix the transformation to BTZ coordinates \eqref{BTZ_metric}. However, there are four equivalent transformations leading to Matsubara modes when expressed in BTZ coordinates
\begin{equation*}
    \left\{\begin{aligned}
        x_-=\pm(r_+\:t -i\: r_-\:\phi)\qquad&\& \qquad x_+=i\:r_-\: t +r_+\phi 
        \\
        x_-=\pm(r_+\:t -i\: r_-\:\phi)\qquad&\& \qquad x_+=-(i\:r_-\: t +r_+\phi)
    \end{aligned}\right.
\end{equation*}
These transformations, however, do not generate new modes; they simply interchange the behavior of the same modes obtained in \eqref{graviton_modes}. Moreover, the first line provides an alternative way to obtain the $n<0$ modes, since changing the sign maps $n>0$ to $-n>0$, both satisfying $\kappa_->0$. }
\begin{equation}\label{eq:axisymmetric_modes}
    e^{i(\kappa_+ x^+ + \kappa_- x^-)}=e^{i\:2\pi nT\:t}\quad\Longrightarrow\quad \kappa_-=n \qquad\&\qquad \kappa_+=-i\: n\frac{r_-}{r_+}
\end{equation}
with $n\in\mathbb{Z}$. Note that we only consider axisymmetric modes; although modes with $\phi$-dependence contribute to the path integral, they do not affect the Schwarzian sector, which dominates near extremality, as we show in Sec.~\ref{sec:normalizable_modes}. Plugging this ansatz into \eqref{graviton_eqns}, the remaining radial equations for $R_{ij}(z)$ take the form of hypergeometric equations. The corresponding solutions can be written as
\begin{equation}\label{graviton_solutions}
    \begin{aligned}
        R_{11}=& e_{11}\:(z-1)^{\frac{1}{2}-\frac{i k}{2}}z^{\frac{|n|}{2}} \, _2F_1\left[a+1,b+1,c,z\right] 
        \\
        R_{12}=& e_{12}\:(z-1)^{-\frac{1}{2}-\frac{i k}{2}} z^{\frac{|n|}2} \,_2F_1\left[a,b,c,z\right] 
        \\
        R_{22}=& e_{22}\:(z-1)^{\frac{3}{2}-\frac{i k}{2}}z^{\frac{|n|}2} \, _2F_1\left[a+2,b+2,c,z\right] 
    \end{aligned}
\end{equation}
where $e_{ij}$ are polarization constants, and we defined the parameters 
\begin{equation}
    a=\frac{1}{2}\left(2\pi\:|n|\frac{T_L}{r_+} -ik-1\right)\:,\qquad b=\frac{1}{2} \left(2\pi\:|n|\frac{T_R}{r_+} -ik -1\right) \:,\qquad c=|n|+1
\end{equation}
with the left/right moving temperatures defined in \eqref{eq:T_LR}. Here we used the fact that each solution consists of two terms that interchanges $n\leftrightarrow-n$, then they have opposite behavior at the horizon: for $n>0$, one converges as $z\to 0$ while the other diverges, and vice versa for $n<0$. In the Lorentzian case one typically selects the ingoing solution at the horizon \cite{Indios}. In the Euclidean case, however, it is less clear which solution is the physical one, and we will therefore keep both branches. Nevertheless, for any fixed sign of $n$, only a single branch is regular at the horizon, thus we retain only this regular part for $n\in\mathbb{Z}$. 

Finally, returning to the original tensor basis via \eqref{graviton_transformation} and imposing that also satisfies the first-order equation \eqref{graviton_linear}, implies that the polarizations constants are not independent. The easiest way to compute them is in the limit $z\to0$ and, since the polarizations are constants, the obtained relation holds for all $z$
\begin{equation}\label{graviton_polarizations}
    \begin{aligned}
        e_{11} &= -\frac{2(2\pi|n|T_L +ik\:r_+ -r_+)}{2\pi|n|T_R -ik\:r_+ +r_+}e_{22} = -\frac{2(2\pi|n|T_R -ik\:r_+ -r_+)}{2\pi|n|T_L +i k\:r_++r_+}e_{12} 
        \\
        e_{12} &=  \frac{(2\pi|n|T_L +ik\:r_+ -r_+)(2\pi|n|T_L +ik\:r_+ +r_+)}{(2\pi|n|T_R -ik\:r_+-r_+) (2\pi|n|T_R-i k\:r_++r_+)}e_{22}
        \\
        e_{22} &= \frac{(2\pi|n|T_R -ik\:r_+ -r_+)(2\pi|n|T_R -ik\:r_++r_+)}{(2\pi|n|T_L+ik\:r_+-r_+)(2\pi|n|T_L +ik\:r_+ +r_+)}e_{12}
    \end{aligned}
\end{equation}
These polarizations correspond to the $+$ sign in \eqref{graviton_linear}, while exchanging $T_R\leftrightarrow T_L$ yields the polarization associated with the opposite sign. Thus, although there are three possible polarizations, they can always be expressed in terms of a single independent constant. We write them all explicitly, since they will be crucial in the search of normalizable modes. 

Notice that the radial part of the scalar solutions \eqref{graviton_solutions} do not depend on the sign of $n$ (including the polarization constants $e_{ij}$), only of its absolute value $R_{ij}^{(|n|)}$. Moreover, the spin-2 transformations \eqref{graviton_transformation} do not depend on $n$; thus, the radial part of the components $R_{++}^{(n)}$, $R_{-+}^{(n)}$ and $R_{--}^{(n)}$ only depend on $|n|$. From \eqref{graviton_zcomponents}, the only radial part components that depend on the sign of $n$ are $R_{z+}$ and $R_{z-}$, which can be written as
\begin{equation}\label{R_zi_n}
    R_{z+}^{(n)}=\frac{n}{|n|}R_{z+}^{(|n|)} \qquad\&\qquad R_{z-}^{(n)}=\frac{n}{|n|}R_{z-}^{(|n|)}
\end{equation}
This will allow us to simplify the search for normalizable modes in the next section.

\subsection{Normalizable modes}\label{sec:normalizable_modes}

We now turn to the analysis of normalizable modes. The solutions are obtained using the Matsubara expansion \eqref{eq:anzat}; therefore, physical perturbations must be constructed as real linear combinations of the complex basis solutions \cite{Joacokerr},
\begin{equation}\label{joaco_modes}
    \tilde{h}_{\mu\nu}=\frac1{\sqrt{2}}\left(h_{\mu\nu}^{(n)} +h_{\mu\nu}^{(-n)}\right)
\end{equation}
with $\tilde{h}^{\mu\nu}\equiv g^{\mu\alpha}g^{\nu\beta}h_{\alpha\beta}$. Normalizability is defined by the ultralocal measure, namely, by the real bilinear form induced by the quadratic expansion of the Euclidean Einstein--Hilbert action \cite{JOACO},
\begin{equation}\label{graviton_unitarity}
    ||h||^2=\int d^3x \:||h||^2_{\sf den}= \int d^3x \:\sqrt{g} \:\tilde{h}_{\mu\nu}\tilde{h}^{\mu\nu}= \int d^3x \:\sqrt{g} \:h_{\mu\nu}^{(n)}h^{\mu\nu\:(-n)}
\end{equation}
where $||h||^2_{\sf den}$ is the norm density of states. The last equality is a consequence of the orthonormality relation for the Matsubara modes with respect to the inner product defined by the Euclidean time integral. Expanding the norm and using \eqref{graviton_zcomponents} and \eqref{R_zi_n} to express it in terms of the independent components of the perturbation, we obtain
\begin{equation}
    \begin{aligned}
        ||h||^2_{\sf den}&=\frac{1}{z^4}\left(a_1\:h_{--}^{(n)}h_{--}^{(-n)} +a_2\:h_{--}^{(n)}h_{-+}^{(-n)} +a_3\:h_{-+}^{(n)}h_{-+}^{(-n)} +a_4\:h_{--}^{(n)}h_{++}^{(-n)} \right.
        \\
        &\hspace{8.5cm}\left.+ a_5\:h_{-+}^{(n)}h_{++}^{(-n)} + a_6\:h_{++}^{(n)}h_{++}^{(-n)}\right)
    \end{aligned}
\end{equation}
where the coefficients $a_i$ are polynomials in $(1-z)$,
\begin{equation}\label{poly}
    \begin{aligned}
        a_1&= 1 +n^2\frac{r_-^2}{r_+^2}\left(-\frac{(1-z)}{\left(k^2+1\right) }+\frac{(1-z)^2}{\left(k^2+1\right)^2}\right)
        \\
        a_2&= 2 i\:n^2\frac{r_-}{r_+}\left(\frac{(1-z)}{(k^2+1)} -\left(1-ik \frac{r_-}{r_+}\right)\frac{(1-z)^2 }{\left(k^2+1\right)^2}\right)
        \\
        a_3&=1 -\left(1+k^2-n^2 \left(1-\frac{r_-^2}{r_+^2}\right) \right)\frac{(1-z)}{\left(k^2+1\right)}-n^2 \left(1-ik \frac{r_-}{r_+}\right)^2\frac{(1-z)^2}{\left(k^2+1\right)^2}
        \\
        a_4&=1-(1-z) +2i\:k\: n^2\frac{r_-}{r_+}\frac{(1-z)^2}{\left(k^2+1\right)^2}
        \\
        a_5&= 2 i\: n^2\frac{r_-}{r_+}\frac{(1-z)}{(k^2+1)} +2\:k\:n^2\left(1-ik\frac{r_-}{r_+}\right)\frac{(1-z)^2}{\left(k^2+1\right)^2}
        \\
        a_6&=1 -\left(2+2k^2-n^2\right)\frac{(1-z)}{k^2+1}  + \left(\left(k^2+1\right)^2-k^2 n^2\right)\frac{(1-z)^2}{\left(k^2+1\right)^2}
    \end{aligned}
\end{equation}
Since each of these components is a combination of smooth functions, determining the convergence of the norm reduces to studying their asymptotic behavior. Hence, we only need to analyze the asymptotic behavior of the combinations of $h_{--}$, $h_{-+}$ and $h_{++}$ given in \eqref{graviton_transformation}. Notice that the solutions \eqref{graviton_solutions} have the form $h_{ij}^{(n)}=e^{i\,2\pi nT\,t}\,R_{ij}^{(|n|)}(z)$, thus the only remaining part is the radial one,
\begin{equation}\label{graviton_norm}
    \begin{aligned}
        h_{--}^{(n)}h_{--}^{(-n)}&= (R_{11})^2 -2R_{11}R_{12}+(R_{12})^2-2R_{11}R_{22}+2R_{12}R_{22} +(R_{22})^2
        \\
        h_{--}^{(n)}h_{-+}^{(-n)}&=i (R_{12})^2+i R_{11} R_{12}-i R_{11} R_{22}-i (R_{22})^2
        \\
        h_{-+}^{(n)}h_{-+}^{(-n)}&= -(R_{12})^2 +2R_{12}R_{22}-(R_{22})^2
        \\
        h_{--}^{(n)}h_{++}^{(-n)}&= (R_{11})^2-2 R_{12} R_{22}-R_{12}^2-R_{22}^2
        \\
        h_{-+}^{(n)}h_{++}^{(-n)}&= i R_{11} R_{12}-i R_{11} R_{22}-i R_{12}^2+i R_{22}^2
        \\
        h_{++}^{(n)}h_{++}^{(-n)}&= (R_{11})^2 +2R_{11}R_{12} +(R_{12})^2 +2R_{11}R_{22}+2R_{12}R_{22}+(R_{22})^2
    \end{aligned}
\end{equation} 
where each $R_{ij}$ depends on $|n|$, although we do not write it explicitly to simplify notation.

We next examine the near-horizon ($z\to0$) and near-boundary ($z\to1$) behaviors. In the near-horizon limit, the modes \eqref{graviton_norm} are regular and approach zero for $|n|>1$ without requiring any additional conditions. On the other hand, near the boundary $(z\to1)$, keeping the leading terms in the polynomials \eqref{poly}, the norm density takes the form
\begin{equation}\label{eq:norm_density}
    \begin{aligned}
        ||h||^2_{\sf den}\sim & \frac{1}{z^2}\left(3 (R_{11})^2 +4\:R_{12}R_{22} +\alpha_+(1-z)^2(R_{12})^2 +\alpha_-(1-z)^2(R_{22})^2 \right.
        \\
        &\left.
        \qquad+\beta_+ (1-z)R_{11}R_{12}  +\beta_{-} (1-z)R_{11}R_{22}
        \right)
    \end{aligned}
\end{equation}
where
\begin{equation}
    \alpha_{\pm}=1+\frac{n^2 \left(1\pm \frac{r_-}{r_+}\right)^2}{(1\pm ik)^2}\:, \qquad \beta_{\pm}=4-\frac{2n^2 \left(1\pm \frac{r_-}{r_+}\right)^2}{\left(k^2+1\right)}
\end{equation}
Hence, to determine normalizability we only have to analyze the asymptotic behavior of the radial functions $R_{ij}(z)$ in \eqref{graviton_solutions}. Using the hypergeometric expansion
\begin{equation}
    \, _2F_1\left[a,b,c,z\right]\sim \frac{\Gamma(c-a-b)\Gamma(c)}{\Gamma(c-a)\Gamma(c-b)} +\frac{\Gamma(a+b-c)\Gamma(c)}{\Gamma(a)\Gamma(b)}(1-z)^{c-a-b} 
    \label{asymp}
\end{equation}
the radial functions behave as
\begin{equation}\label{graviton_scalars_asym}
    \begin{aligned}
        R_{11}\sim\:& e_{11} (z-1)^{\frac{1}{2}-\frac{i k}{2}}\left(\frac{\Gamma(c-a-b-2)\Gamma(c)}{\Gamma(c-a-1)\Gamma(c-b-1)} +\frac{\Gamma(a+b+2-c)\Gamma(c)}{\Gamma(a+1)\Gamma(b+1)}(1-z)^{i k} \right) 
        \\
        R_{12}\sim\:& e_{12} (z-1)^{-\frac{1}{2}-\frac{i k}{2}}\left(\frac{\Gamma(c-a-b)\Gamma(c)}{\Gamma(c-a)\Gamma(c-b)} +\frac{\Gamma(a+b-c)\Gamma(c)}{\Gamma(a)\Gamma(b)}(1-z)^{i k+2}\right) 
        \\
        R_{22}\sim\:& e_{22} (z-1)^{\frac{3}{2}-\frac{i k}{2}}\left(\frac{\Gamma(c-a-b-4)\Gamma(c)}{\Gamma(c-a-2)\Gamma(c-b-2)} +\frac{\Gamma(a+b+4-c)\Gamma(c)}{\Gamma(a+2)\Gamma(b+2)}(1-z)^{i k-2} \right)
    \end{aligned}
\end{equation}
Inserting these asymptotic behaviors into the norm density \eqref{eq:norm_density}, normalizable modes require that the leading term decays faster than $\mathcal{O}((1-z)^{-1})$ near the boundary. In particular, a logarithmic divergence arises from
\begin{equation}\label{log_divergence}
    \begin{aligned}
        R_{12}\:R_{22}\sim e_{12}\:e_{22}\:\frac{\Gamma (c)^2 \Gamma (a+b-c+4) \Gamma (-a-b+c)}{\Gamma (a+2) \Gamma (b+2) \Gamma (c-a) \Gamma (c-b)}\:(1-z)^{-1}
    \end{aligned}
\end{equation}
Notice that this divergent term is independent of $k$, and it is the only term of this form in \eqref{eq:norm_density}. In principle, there are two ways to cancel it: (\textit{i}) set either of the polarizations $e_{12}$ or $e_{22}$ to zero; or (\textit{ii}) impose that the corresponding coefficient involving Gamma functions vanishes. The second possibility is the standard mechanism discussed in the Lorentzian literature \cite{Indios}. We will show, however, that in the present Euclidean case only the first option leads to genuinely normalizable modes.

Normalizable modes contributing to the gravitational path integral must satisfy the condition $1+k^2\in \mathbb{R}$, so $k$ is either real or purely imaginary. Notice that both possibilities (\textit{i}) and (\textit{ii}) involve a constraint if fulfilled for $ik\in\mathbb{R}$, but if $k\in\mathbb{R}$, the only option is to trivially vanish the polarizations in (\textit{i}). Therefore, we cannot have simultaneously the real $k$ case and satisfy the ultralocal measure \eqref{graviton_unitarity}\footnote{Real modes are also discussed in~\cite{Leston}, where the only normalizable modes are found to have $k\in\mathbb{R}$. This does not contradict our result, since they search for unitary modes using the standard $\mathcal{L}_2$ inner product, $\sqrt{g}\,h_{\mu\nu}^*h^{\mu\nu}$, in which case the logarithmic divergence \eqref{log_divergence} cancel out (additionally, they perform the rotation $r_-\rightarrow i\,r_-$). In contrast, we require modes to satisfy the ultralocal measure \eqref{graviton_unitarity} since these modes are the on. Moreover, real $k$ never leads to zero modes of the Lichnerowicz operator \eqref{graviton_lich}, since $k^2+1>0$ for all $k\in\mathbb{R}$. We therefore conclude that the only normalizable modes satisfying the norm \eqref{graviton_unitarity} are the purely imaginary modes studied here.}. Thus we focus on purely imaginary modes.

\paragraph{(\textit{i}) Vanishing polarizations.} As mentioned above, normalizability requires the leading term in \eqref{eq:norm_density} to decay faster than $(1-z)^{-1}$. In particular, setting the polarizations $e_{12}$ or $e_{22}$ to zero using \eqref{graviton_polarizations} fixes the dependence of $k$ on the temperature.

Let us consider first the case with $e_{12}=0$; then the norm density \eqref{eq:norm_density} becomes
\begin{equation}
    \begin{aligned}
        ||h||^2_{\sf den}\sim & \frac{1}{z^2}\left(3 (R_{11})^2 +\alpha_-(1-z)^2(R_{22})^2 +\beta_{-} (1-z)R_{11}R_{22}
        \right)
    \end{aligned}
\end{equation}
From the asymptotic expansion \eqref{graviton_scalars_asym}, each of these terms has a convergence range required for normalizability,
\begin{table}[h!]
    \centering
    \begin{tabular}{ll}
        $(R_{12})^2$ & is normalizable if $-2\leq ik\leq2$ \\
        $(1-z)\:R_{11}\:R_{22}$ & is normalizable if $-2\leq ik\leq4$   \\
        $(1-z)^2\:(R_{22})^2$ & is normalizable if $-2\leq ik\leq6$
    \end{tabular}
\end{table}
\\
Notice that all these terms are normalizable for $-2\leq ik\leq2$. On the other hand, from the polarizations \eqref{graviton_polarizations}, notice that there are two choices to satisfy $e_{12}=0$,
\begin{equation}\label{graviton_modes}
    k_{\text{sch}}=-i+i\:|n|\left(1-\frac{r_-}{r_+}\right) \qquad \text{or}\qquad k_{\sf rot}=i+i\:|n|\left(1-\frac{r_-}{r_+}\right)
\end{equation}
where the subscripts denote the Schwarzian mode and rotational mode, as will be clear in the next sections\footnote{We simplify the computation by focusing on axisymmetric modes and using the transformation \eqref{eq:axisymmetric_modes}. The general solution for the eigenvalue problem with the ansatz \eqref{eq:anzat} is $k=\kappa_++i\,\kappa_-\pm i$. The transformations \eqref{btz_transformations} then yield the expansion in terms of the BTZ coordinates \eqref{BTZ_metric}. If we were to consider non-axisymmetric modes, we would modify the Matsubara expansion to $e^{i\,2\pi\,T(n\,t+\omega\,\phi)}$, with $\omega\in\mathbb{R}$ to obtain oscillatory modes. This expansion leads to
\begin{equation*}
    \kappa_-=n-i\,\omega\frac{r_-}{r_+} \qquad\&\qquad \kappa_+=-i\,n\frac{r_-}{r_+}-\omega \qquad\Longrightarrow\qquad k= i\,|n| \left(1-\frac{r_-}{r_+}\right)\pm i-\omega  \left(1-\frac{r_-}{r_+}\right)
\end{equation*}
We are searching for modes of the Lichnerowicz operator \eqref{graviton_lich} that can become zero modes, which requires $k$ to be purely imaginary. Therefore, there are no Schwarzian modes for real and non-zero $\omega$. Consequently, while non-axisymmetric modes contribute to the one-loop correction the of path integral in Sec.~\ref{sec:1loop}, they do not belong to the Schwarzian sector of interest.}. In particular, notice from \eqref{graviton_polarizations} that the $k_{\text{sch}}$ condition not only makes $e_{12}$ vanish, but also implies $e_{11}=0$. Thus, this mode only leaves one scalar field ($h_{22}$), and the corresponding perturbation takes the form
\begin{equation}\label{graviton_schwarzian_z}
    h_{\mu\nu}^{(n)}(k_{\text{sch}})\:dx^\mu dx^\nu=  \frac{\mathcal{N}_n}4\:e^{i\:2\pi nT\:t}\:(1-z)^{\frac{|n|}{2}  
   \left(\frac{r_-}{r_+}-1\right)}\:z^{\frac{|n|}2-2}\: (dz+2 (r_--r_+)\:z\:(d\phi-i dt))^2
\end{equation}
Integrating these modes we obtain the norm \eqref{graviton_unitarity}
\begin{equation}
    ||h||^2 = \frac{8 \pi ^2 \:(\mathcal{N}_{|n|})^{2}\:r_+\: \Gamma (|n|-1) \:\Gamma \left(3+|n|\left(\frac{r_-}{r_+}-1\right)\right)}{\Gamma \left(2+|n|\frac{r_-}{r_+}\right)},\qquad 1<|n|<\frac{3\:r_+}{r_+-r_-}
\end{equation}
Thus, we can define the normalization as 
\begin{equation}\label{schwarzian_norm}
    (\mathcal{N}_{|n|})^{2}= \frac{\Gamma \left(2+|n|\frac{r_-}{r_+}\right)}{8 \pi ^2 \:r_+\: \Gamma (|n|-1) \:\Gamma \left(3+|n|\left(\frac{r_-}{r_+}-1\right)\right)} 
\end{equation}
We will discuss the properties and physical interpretation of these normalizable modes in Sec.~\ref{sec:Schwarzian}.

We now compute the other mode of interest obtained with $k_{\sf rot}$. This mode involves a linear combination of the scalars $h_{11}$ and $h_{22}$,
\begin{equation}\label{graviton_rotational_z}
    \begin{aligned}
        &h_{\mu\nu}^{(n)}(k_{\sf rot})= \tilde{\mathcal{N}}_n\:e^{i\:2\pi nT\:t}\:\frac{z^{\frac{|n|}{2}-2} (1-z)^{\frac{1}{2} |n| \left(\frac{r_-}{r_+}-1\right)-1} }{4\:r_+\: (|n|+1)}(dz+2 z \:\frac{n}{|n|}(i\:dt-d\phi) (r_+-r_-))\times
        \\
        &\times\left((z+1)\:dz+2 i \frac{n}{|n|}z (r_- (3-z)+r_+ (1-3 z))\:dt +2 \frac{n}{|n|} z (r_-(1-3z)+r_+(3-z))\:d\phi\right.
        \\
        & \qquad\qquad\left.-\frac{|n|}{r_+} (r_+-r_- z) (dz+2 z \frac{n}{|n|}(i\:dt-d\phi) (r_+-r_-))\right)
    \end{aligned}
\end{equation}
Integrating these modes we obtain the norm \eqref{graviton_unitarity}
\begin{equation}\label{rotational_norm}
    ||h||^2 = \frac{8 \pi^2 \:(\tilde{\mathcal{N}}_{|n|})^{2} \left(|n| (r_--r_+)-2r_+\right)\:\Gamma(|n|)\:\Gamma\left(|n| \left(\frac{r_-}{r_+}-1\right)+1\right)}{(|n|+1)^2\:\Gamma\left(|n|\frac{r_-}{r_+}+1\right)},\quad\:\: 1<|n|<\frac{r_+}{r_+-r_-}
\end{equation}
Thus, we can define the normalization as 
\begin{equation}
    (\tilde{\mathcal{N}}_{|n|})^{2}= \frac{(|n|+1)^2\:\Gamma\left(|n|\frac{r_-}{r_+}+1\right)}{8 \pi^2 \: \left(|n| (r_--r_+)-2r_+\right)\:\Gamma(|n|)\:\Gamma\left(|n| \left(\frac{r_-}{r_+}-1\right)+1\right)}
\end{equation}
This mode has the features expected for the rotational mode, which will be discussed in Sec.~\ref{sec:rotational_modes}.

The previous result corresponds to taking $e_{12}=0$ in order to cancel the logarithmically divergent term \eqref{log_divergence}. Alternatively, one could use the polarizations \eqref{graviton_polarizations} to take $e_{22}=0$ instead. This yields normalizable modes, but they will not give zero modes in the extremal case, which is what we are looking for. However, we now analyze these modes to exhaust all normalizable modes satisfying the ultralocal measure \eqref{graviton_unitarity}. Taking $e_{22}=0$, the norm density \eqref{eq:norm_density} becomes  
\begin{equation}
    \begin{aligned}
        ||h||^2_{\sf den}\sim & \frac{1}{z^2}\left(3 (R_{11})^2 +\alpha_+(1-z)^2(R_{12})^2 +\beta_+ (1-z)R_{11}R_{12}
        \right)
    \end{aligned}
\end{equation}
From the asymptotic expansion \eqref{graviton_scalars_asym}, each of these terms has a convergence range required for normalizability, 
\begin{table}[H]
    \centering
    \begin{tabular}{ll}
        $(R_{11})^2$ & normalizable if $-2\leq ik\leq2$ \\
        $(1-z)\:R_{11}\:R_{12}$ & normalizable if $-4\leq ik\leq2$   \\
        $(1-z)^2\:(R_{12})^2$ & normalizable if $-6\leq ik\leq2$
    \end{tabular}
\end{table}
Notice that all these terms are normalizable for $-2\leq ik\leq2$. On the other hand, from the polarizations \eqref{graviton_polarizations}, notice that there are two choices to satisfy $e_{22}=0$,
\begin{equation}\label{non_zero_spectrum}
    \tilde{k}_{\sf n.z.}=-i-i\:|n|\left(1+\frac{r_-}{r_+}\right) \qquad \text{or}\qquad k_{\sf n.z.}=i-i\:|n|\left(1+\frac{r_-}{r_+}\right)
\end{equation}
where the subscript ``nz'' denotes non-zero, indicating that these modes do not become zero modes in the extremal limit. Recalling that $|n|\geq2$, we observe that $i\tilde{k}_{\text{nz}}<-2$; therefore, this mode is non-normalizable. Thus, the only normalizable mode is given by $k_{\sf nz}$, which is only proportional to the scalar $h_{12}$. For completeness, and to facilitate the discussion in Sec.~\ref{sec:Kerr_Schild}, we also record here the explicit expression for this perturbation
\begin{equation}\label{graviton_nonschwarzian}
    h_{\mu\nu}^{(n)}(k_{\sf n.z.})\:dx^\mu dx^\nu =\mathcal{N}_{\sf n.z.}\:e^{i\:2\pi nT\:t}\: z^{\frac{|n|}{2}-2} (1-z)^{-\frac{|n|}2 \left(\frac{r_-}{r_+}+1\right)} \left(\frac{dz}2+\frac{n}{|n|} z (r_-+r_+)(i\: dt+d\phi)\right)^2
\end{equation}
Integrating these modes we obtain the norm \eqref{graviton_unitarity}
\begin{equation}
    ||h||^2 = \frac{8 \pi ^2 (\mathcal{N}_{\sf n.z.})^{2}\:r_+\:\Gamma(|n|-1)\:\Gamma\left(3-|n|\left(\frac{r_-}{r_+}+1\right)\right)}{\Gamma \left(2-|n|\frac{r_-}{r_+}\right)} ,\qquad 1<|n|<\frac{3\:r_+}{r_++r_-}
\end{equation}
Thus, we can define the normalization as
\begin{equation}
    (\mathcal{N}_{\sf n.z.})^{2}= \frac{\Gamma \left(2-|n|\frac{r_-}{r_+}\right)}{8 \pi^2\:r_+ \:\Gamma(|n|-1) \Gamma \left(3-|n|\left(\frac{r_-}{r_+}+1\right)\right)} 
\end{equation}
These are normalizable modes; however, we are interested in modes that become zero modes of the Lichnerowicz operator \eqref{graviton_lich} in the extremal limit $r_-\rightarrow r_+$, i.e., $k^2=-1$, which $k_{n.z.}$ does not satisfy.

\paragraph{(\textit{ii}) Vanishing coefficients via Gamma functions.} Alternatively, one might attempt to obtain normalizable modes by forcing the coefficients in \eqref{log_divergence} to vanish, using the poles of the Gamma functions. That is, by requiring the argument to be a negative integer, $\Gamma(-l)^{-1}=0$ with $l\in\mathbb{Z}_+$. From \eqref{log_divergence}, we have only four possibilities to consider,
\begin{equation}
    ik=\pm\left(3+2l+|n|\left(1-\frac{r_-}{r_+}\right)\right) \qquad\&\qquad ik=\pm\left(3+2l+|n|\left(1+\frac{r_-}{r_+}\right)\right)
\end{equation}
However, these relations lead to divergent branches in the asymptotic expansions \eqref{eq:norm_density} that do not cancel. Thus, none of these relations can be satisfied. We therefore conclude that normalizable modes cannot be obtained by tuning the Gamma function coefficients. The only admissible normal modes are those found by setting the polarizations to zero, as given in \eqref{graviton_schwarzian_z}, \eqref{graviton_rotational_z} and \eqref{graviton_nonschwarzian}.

\newpage

\section{Modes contributing to the Logarithmic Correction}\label{sec:modes}
In the previous section we showed that, upon fixing a sign convention in the linear equation \eqref{graviton_linear}, there exist two normalizable modes that become zero modes in the extremal limit. These are simultaneous eigenfunctions of the first-order operator and the Lichnerowicz operator, satisfying
\begin{equation}\label{Lichnerowichz_spectrum}
\epsilon_\mu^{\:\:\alpha\beta}\nabla_\alpha h_{\beta\nu} = k_n\: h^{(n)}_{\mu\nu}\:,\qquad \nabla_{L}\:h_{\mu\nu}^{(n)}=\alpha_n\:h_{\mu\nu}^{(n)}
\end{equation}
where the possible values of $k_n$ were obtained for this sign convention in \eqref{graviton_modes} (the opposite sign choice yields the complex conjugate), and we have defined the Lichnerowicz spectrum as $\alpha_n\equiv 1+k_{n}^2$. In this section, we first analyze the physical properties of these modes and then compute their contribution to the gravitational path integral.

\subsection{Schwarzian modes}\label{sec:Schwarzian}
First, let us analyze the modes corresponding to $k_{\text{sch}}$ in \eqref{graviton_modes}. The Lichnerowicz spectrum in this case is given by
\begin{equation}\label{schwarzian_spectrum}
    \alpha_{\text{sch}}^{(n)} = 1-\left(1+|n|\left(\frac{r_-}{r_+}-1\right)\right)^2 \:,\qquad 1<|n|<\frac{3\:r_+}{r_+-r_-} 
\end{equation}
A notable property of the modes obtained in \eqref{graviton_schwarzian_z} is that they admit a simple factorized form,
\begin{equation}\label{graviton_schwarzian_r}
    h_{\mu\nu}^{(n)}(k_{\text{sch}})=\Phi^{(n)}(t,r)\:K_\mu K_\nu 
\end{equation}
where we have used the transformations \eqref{btz_transformations} to express the result in BTZ coordinates \eqref{BTZ_metric}. The scalar field is defined as
\begin{equation}\label{scalar_graviton}
       \Phi_n(t,r)= \mathcal{N}_{|n|}\:e^{i\:2\pi nT\:t}\:\left(r^2-r_+^2\right)^{\frac{|n|}{2}} \left(r_+^2-r_-^2\right)^{-\frac{|n|}{2} \left(1-\frac{r_-}{r_+}\right)}    \left(r^2-r_-^2\right)^{-\frac{|n|}{2}\frac{r_-}{r_+}}(r_--r_+)^2
\end{equation}
and the one-form reads
\begin{equation}\label{graviton_factorize}
    K_\mu dx^\mu= \frac{(r_- +r_+)}{r\:f(r)} \:dr+\frac{n}{|n|}(i \:dt  -d\phi)
\end{equation}
where $f(r)$ is the lapse function defined in \eqref{BTZ_metric}. This vector field appears as a purely geometric feature, and it is traceless and divergence-free\footnote{The modes \eqref{graviton_factorize} are complex, and there is no combination of fields that can provide a real mode. However, a Wick rotation $t\to i t$ yields a real field.}. We can also point out that $\Phi^{(n)}(t,r) K_\mu$ is exactly the Schwarzian mode found in appendix \ref{sec:vector} for the vector field. This relation between gravitational perturbations and gauge theories has been noticed previously and motivates the so-called double copy theory \cite{Bern:2010}. In the specific context of three-dimensional gravity, a similar construction was explored in \cite{Carrillo:2019}, where the BTZ black hole was interpreted as a classical double copy. In our case, however, the background metric is already BTZ, and the perturbations themselves behave as gauge fields around this background. We will expand on this point in Sec.~\ref{sec:Kerr_Schild}.

Although the modes in \eqref{graviton_schwarzian_r} are normalizable with respect to the ultralocal measure \eqref{graviton_norm} within the range $1<|n|<3 r_+/(r_+-r_-)$, physical consistency imposes stricter constraints. We require the perturbations to preserve the asymptotic structure of the spacetime; that is, the total metric must remain asymptotically AdS$_3$ as $r\to\infty$. Consider the asymptotic behavior of the Schwarzian modes,
\begin{equation}\label{bcc}
\begin{aligned}
    h_{tt} &\propto r^{m}  \qquad\qquad&& h_{\phi r} \propto r^{-3+m} \\
    h_{tr} &\propto r^{-3+m} && h_{t\phi} \propto r^{m} \\
    h_{rr} &\propto r^{-6+m}  && h_{\phi \phi} \propto r^{m} \\
\end{aligned}
\end{equation}
where we have defined the exponent $m=|n|(1-(r_-/r_+))$, thus, $0\leq m<3$. Comparing this with the background BTZ metric \eqref{BTZ_metric}, where for instance $g_{tt}\propto r^2$, it is evident that we must impose $m<2$. Modes with $m\geq2$ would produce perturbations that decay slower than (or are comparable to) the background, thereby modifying the asymptotic AdS$_3$ boundary conditions. On the other hand, if $0\leq m<2$, all components remain subleading relative to the background metric. Consequently, the spectrum of normalizable modes satisfying physically admissible boundary conditions is restricted to
\begin{equation}\label{schwarzian_spectrum_2}
    \alpha_{\text{sch}}^{(n)} = 1-\left(1+|n|\left(\frac{r_-}{r_+}-1\right)\right)^2 \:,\qquad 1<|n|<\frac{2\:r_+}{r_+-r_-} 
\end{equation}
It is important to note that these modes generally do not satisfy the standard Brown-Henneaux boundary conditions \cite{BH_bc}, which, in particular, require $h_{t\phi}\sim\mathcal{O}(1)$. Since $h_{t\phi}\propto r^m$ with $m\geq0$, these Schwarzian modes do not preserve the conventional conformal symmetry on the boundary (they are only preserved in the extremal case with $m=0$). Instead, they satisfy generalized boundary conditions (see, e.g., \cite{Grumiller,CSS_2013}) which allow for dynamical fluctuations of the boundary metric. In the holographic picture, this corresponds to turning on a source at the boundary while maintaining the asymptotically AdS$_3$ character of the bulk.

\begin{figure}
	\begin{center}
	\includegraphics[width=0.5\textwidth]{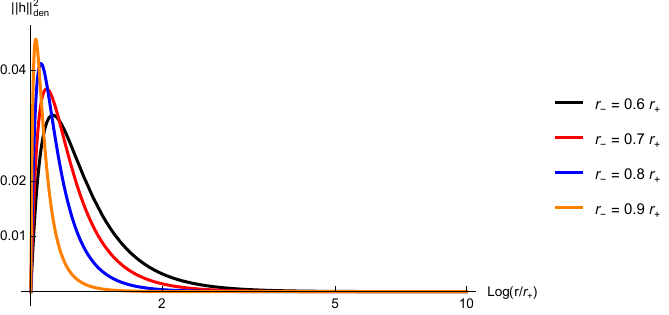}%
	\includegraphics[width=0.5\textwidth]{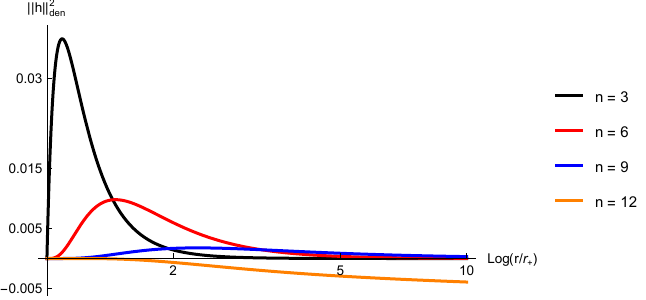}
    \caption{Norm density of the Schwarzian modes, $||h||^2_{\sf den}=\sqrt{g}\,h_{\mu\nu}^{(n)}h^{\mu\nu\,(-n)}$, as a function of the proper radial coordinate $\log(r/r_+)$. \textbf{Left}: Fixed mode $n=3$ for different ratios $r_-/r_+$, this plot exactly matches the one found in \cite{JOACO}, confirming that we have recovered the previous results. The modes become increasingly localized near the horizon as extremality is approached ($r_- \to r_+$), eventually concentrating at the horizon. Away from extremality, the modes spread out and eventually become non-normalizable for higher $n$, as predicted by \eqref{schwarzian_spectrum}, which manifests as a negative norm density as can be seen in the right plot. \textbf{Right}: Norm density for several modes at fixed $r_-=0.7\,r_+$. In agreement with \eqref{schwarzian_spectrum}, only a finite number of Schwarzian modes are normalizable; in this case, modes with $n \geq 10$ become non-normalizable, while the lowest modes are increasingly localized near the horizon.}
    \label{fig:Schwarzians}
	\end{center}
\end{figure}

We have thus identified in \eqref{schwarzian_spectrum_2} the subset of the Lichnerowicz spectrum that is both normalizable and consistent with the asymptotic geometry. A striking feature of this result is that the number of normal modes is finite for generic black hole parameters. In the rotationless limit ($r_-\to 0$), there is a single normal mode, whereas in the extremal limit ($r_+\to r_-$), an infinite tower of normal modes emerges. This behavior is illustrated in Fig.~\ref{fig:Schwarzians}, where we plot the normalized norm density $||h||^2_{\sf den}$ defined in \eqref{graviton_unitarity}. The emergence of infinitely many normal modes is expected, since in this limit the BTZ geometry develops an $AdS_2\times S^1$ throat, whose isometry group $SL(2,\mathbb{R})\times U(1)$ admits infinitely many Schwarzian modes associated with the $AdS_2$ boundary \cite{Turiaci:2020}. When the temperature is turned on, the number of these modes becomes finite, decreasing as the temperature increases. This behavior is well understood in the near-extremal limit, where the near-horizon geometry becomes $nAdS_2\times S^1$ \cite{Castro:2018,Moitra:2019}. In this low-temperature regime, the Schwarzian modes acquire dynamics described by Jackiw–Teitelboim gravity, with the temperature acting as a regulator for the infinite zero modes.


All these results have been primarily restricted to the near-horizon region of black holes. However, we find that the modes \eqref{graviton_schwarzian_r} actually extend over the full BTZ geometry, reducing to the usual Schwarzian modes only when restricted to the throat. Notably, in the near-extremal limit, the eigenvalues of the Lichnerowicz operator become proportional to the Matsubara frequencies $n$, specifically $\alpha_n\approx 2|n|(r_+-r_-)/r_+$, matching the result obtained in the near-horizon case \cite{Toldo:2024} (we explicitly compute the partition function of these modes in Sec.~\ref{sec:1loop}). For this reason, we refer to \eqref{graviton_schwarzian_r} as ``Schwarzian modes'', even though they are defined on the entire geometry\footnote{Notice that the combination $\tilde{h}_{\mu\nu}(k_{\text{sch}})=\left(h^{(n)}_{\mu\nu}(k_{\text{sch}})+h^{(-n)}_{\mu \nu}(k_{\text{sch}})\right)/\sqrt{2}$, introduced in \eqref{joaco_modes}, reproduces the Schwarzian mode identified in \cite{JOACO}.}. 

\paragraph{Near horizon near extremal limit.} We now show that the Schwarzian modes \eqref{graviton_schwarzian_r} precisely correspond to the modes obtained in the near horizon near extremal limit limit of the BTZ black hole. In the near extremal regime, we expand in powers of the temperature $T$ while keeping the angular momentum $J$ fixed, namely,
\begin{equation}
    r_+=r_0+\frac\pi2 T +\mathcal{O}(T^2)\:,\qquad r_-=r_0-\frac\pi2 T +\mathcal{O}(T^2)
\end{equation}
Expanding the line element \eqref{BTZ_metric}, together with the transformations 
\begin{equation}
    \begin{aligned}
        r&=r_++T\:\pi\: (z-1) \\
        t&=\frac{1}{2 \pi T}\:\tau \\
        \phi&=\theta -i\tau\left(\frac{1}{2 \pi T}-\frac{1}{2 r_+ r_-}\right)
    \end{aligned}
\end{equation}
the metric in the extremal case, takes the form
\begin{equation}
    ds^2_{\text{throat}} = \frac14\left((z^2-1)\:d\tau^2 +\frac{dz^2}{z^2-1} +4r_0^2\left(d\phi-i\frac{z}{2r_0}\:d\tau\right)^2\right)
\end{equation}
which is a fibration of a circle over $AdS_2$. Taking $z=\cosh\eta$, the obtained modes in the full geometry \eqref{graviton_schwarzian_r} reduce in this limit to
\begin{equation}\label{schwarian_NHNE}
    h^{(n)}_{\mu \nu}=4\,\mathcal{N}_n\,e^{i n t}\,\tanh^{|n|}\left(\frac{\eta}{2}\right)\,\left( d\tau+\frac{n}{|n|}\frac{i}{\sinh \eta}\:d\eta \right)^2
\end{equation}
which coincide with the zero modes found in the near horizon near extremal limit analysis of BTZ in \cite{Toldo:2024}.  

\subsection{Rotational modes}\label{sec:rotational_modes}
\begin{figure}
	\begin{center}
	\includegraphics[width=0.5\textwidth]{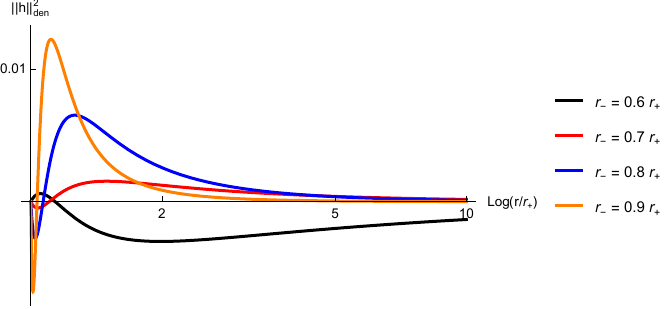}%
	\includegraphics[width=0.5\textwidth]{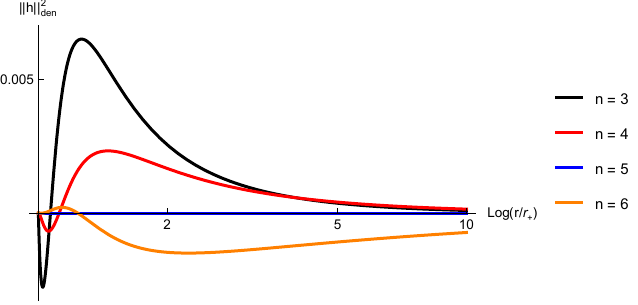}
    \caption{Norm density of the rotational modes, $||h||^2_{\sf den}=\sqrt{g}\,h_{\mu\nu}^{(n)}h^{\mu\nu\,(-n)}$, as a function of the proper radial coordinate $\log(r/r_+)$. \textbf{Left}: Fixed mode $n=3$ for different ratios $r_-/r_+$, with values matching those used in Fig.~\ref{fig:Schwarzians}. The modes become increasingly localized near the horizon as extremality is approached ($r_- \to r_+$), eventually concentrating at the horizon. These modes exhibit negative density near the horizon, but the norm remains well-defined (the integral remains positive). Away from extremality, the modes spread out and eventually become non-normalizable for higher $n$, as predicted by \eqref{rotational_spectrum}, which manifests as a negative norm as seen for $r_-=0.6\,r_+$ (black line). \textbf{Right}: Norm density for several modes at fixed $r_-=0.8\,r_+$. In agreement with \eqref{rotational_spectrum}, only a finite number of rotational modes are normalizable; in this case, modes with $n \geq 5$ become non-normalizable, while the lowest modes are increasingly localized near the horizon.}
    \label{fig:Rotationals}
	\end{center}
\end{figure}

Let us now analyze the other set of modes corresponding to $k_{\text{rot}}$ found in \eqref{graviton_modes}. The Lichnerowicz spectrum \eqref{Lichnerowichz_spectrum} in this case is given by
\begin{equation}\label{rotational_spectrum}
    \alpha_{\text{rot}}^{(n)} = 1-\left(-1+|n|\left(\frac{r_-}{r_+}-1\right)\right)^2 \:,\qquad 1<|n|<\frac{\:r_+}{r_+-r_-} 
\end{equation}
We explicitly obtained the corresponding modes in \eqref{graviton_rotational_z}. Using the transformations \eqref{btz_transformations}, these can also be factorized in BTZ coordinates \eqref{BTZ_metric} as
\begin{equation}
    h^{(n)}_{\mu \nu}(k_{\text{rot}})= \left(\nabla_{\mu}A_\nu+ S_{\mu}A_{\nu} \right) \,dx^\mu dx^\nu
    \label{graviton_rotational_r}
\end{equation}
where $A_{\mu}=\Phi_n(t,r)K_\mu$ is defined in \eqref{graviton_factorize} and corresponds exactly to the Schwarzian mode obtained for the vector field in Appendix~\ref{sec:vector}. Additionally, we have defined the one-form
\begin{equation}
    S_\mu dx^\mu=-\frac{n}{|n|}\frac{(1+i k_{\text{rot}})}{(1+|n|)}\frac{2\pi\, r_+}{T}\left(\left(r^2-r_-^2-r_-r_+ -r_+^2\right)i\, dt-\left(r^2 +r_- r_+\right)\,d\phi\right)
\end{equation}
As in the Schwarzian case, the number of normal modes $\alpha_n$ is finite: there are no normal modes for $r_-<r_+/2$, while an infinite tower of normal modes appears in the extremal limit $r_+=r_-$. However, these modes are tachyonic, in the sense that the eigenvalue of the quadratic operator is always positive; nevertheless, the operator is still bounded. These behaviors are illustrated in Fig.~\ref{fig:Rotationals}, where we plot the normalized norm density $||h||^2_{\sf den}$ defined in \eqref{graviton_unitarity}. Analogously to the Schwarzian case, the modes localize at the horizon as we approach extremality; however, the density becomes negative near the horizon, while the norm (the integral of the density) remains positive.

Notice that the expression \eqref{graviton_rotational_r} resembles a diffeomorphism plus an extra term $S_\mu A_\nu$. This extra term ensures that the fluctuation is indeed a physical mode, since $S^\mu$ is orthogonal to and parallel-transported along the null geodesic vector $A^\mu$,
\begin{equation}
    \begin{aligned}
        S^\mu S_\mu&=\frac{(ik_{\text{rot}}+3)^2}{(|n|+1)^2 (r_+ -r_-)^2} \qquad\qquad&& S^\mu \nabla_\mu A_\nu=\left(2+ik_{\text{rot}}\right)A_\nu\\
        S_\mu A^\mu&=0 \qquad\qquad&& A^\mu \nabla_\mu S_\nu=A_\nu\\
    \end{aligned}
\end{equation}
These properties imply that $S^\mu$ is a vector in the tangent space of BTZ; thus, the obtained mode \eqref{graviton_rotational_r} is not a pure diffeomorphism. Moreover, in the Kerr-Schild language expanded for the Schwarzian modes in Sec.~\ref{sec:Kerr_Schild}, these cross-term perturbations between the transverse vector $S^\mu$ and the null vector $A^\mu$ are common when considering rotational perturbations \cite{Taub:1981}, and are also interpreted as spin perturbations from the gauge point of view \cite{Monteiro:2014}. In this sense, we refer to the perturbation \eqref{graviton_rotational_r} as ``rotational modes''. 

Although the modes in \eqref{graviton_rotational_r} are normalizable with respect to the ultralocal measure \eqref{graviton_unitarity}, we require the perturbations to preserve the asymptotic structure of the spacetime (as discussed for the Schwarzian modes in the previous section); that is, the total metric must remain asymptotically AdS$_3$ as $r\to\infty$. Consider the asymptotic behavior of the rotational modes,
\begin{equation}
\begin{aligned}
    h_{tt} &\propto r^{2+m}  \qquad\qquad&& h_{\phi r} \propto r^{-1+m} \\
    h_{tr} &\propto r^{-1+m} && h_{t\phi} \propto r^{2+m} \\
    h_{rr} &\propto r^{-4+m}  && h_{\phi \phi} \propto r^{2+m} \\
\end{aligned}
\end{equation}
where we have defined the exponent $m=|n|(1-(r_-/r_+))$ as in \eqref{bcc}; note that in this case, $0\leq m<1$. Comparing this with the background BTZ metric \eqref{BTZ_metric}, where for instance $g_{tt}\propto r^2$, we observe that these modes grows faster than the background, or has the same behavior for $m=0$ (the extremal limit), where these modes degenerate into pure diffeomorphisms. The fact that rotational modes might exist in the near-horizon approximation but are ill-defined from the full geometry perspective was also suggested in \cite{JOACO}. We thus conclude that, by explicitly computing the rotational modes in the full BTZ geometry, these modes fail to respect the required asymptotic behavior (unlike the Schwarzian modes \eqref{bcc}), and for this reason, we will not include them in the path integral computation.

\newpage

\subsection{One loop correction from Schwarzian modes}\label{sec:1loop}
Having determined the spectrum of the Schwarzian modes, we now proceed to evaluate their contribution to the one-loop partition function. In the semiclassical approximation, the partition function is expanded around the background saddle-point metric $g_{\mu\nu}$ as $Z \approx e^{-I_{\text{cl}}} Z_{\text{1-loop}}$. The one-loop correction is captured by the path integral over the quadratic fluctuations $h_{\mu\nu}$, determined by the second-order expansion of the Euclidean action $I^{(2)}$
\begin{equation}\label{eq:path_integral_def}
    Z_{\text{1-loop}} = \int \mathcal{D}h_{\mu\nu} \, e^{- \frac{1}{32 \pi G_N} \int d^3x\, \sqrt{g} \, h^{\mu\nu} \Delta_{L} h_{\mu \nu}}
\end{equation}
where $\Delta_{L}$ denotes the Lichnerowicz operator acting on symmetric rank-2 tensors. For three-dimensional gravity in the de Donder gauge ($\nabla^\mu h_{\mu\nu} - \frac{1}{2}\nabla_\nu h = 0$), this operator reduces to the expression given in \eqref{graviton_lich}.

To properly define the path integral, we employ the standard Faddeev-Popov procedure to account for gauge redundancies. The measure splits into contributions from the gauge-fixing term ($GF$), the ghost fields, and the physical transverse-traceless ($TT$) fluctuations. The second-order expansion of the Euclidean action factorizes as
\begin{equation}
    I^{(2)}=I_{GF}+I_{\text{ghost}}+ I_{TT}
\end{equation}
where 
\begin{equation*}
    I_{GF}=\frac{1}{32 \pi G_N} \int dx^3 \sqrt{g} \ \mathcal{G}_\mu \mathcal{G}^\mu
\end{equation*}
and 
\begin{equation*}
    I_{\text{ghost}}=\frac{1}{32 \pi G_N} \int dx^3 \sqrt{g} \ \eta_\mu(-\Box+2)\eta^\mu
\end{equation*}
with the gauge fixing condition $\mathcal{G}_\nu=\nabla^\mu h_{\mu \nu}-\frac12 \nabla_\nu h$, which implies $\delta \mathcal{G}_\mu=(\Box-2) \eta_\mu$. The ghost contribution, arising from the vector fluctuations $\eta^\mu$ associated with infinitesimal diffeomorphisms, is given by the vector Laplacian determinant, $\det(-\Box+2)$. The contribution from these pure gauge modes and trace modes typically cancels or is treated separately depending on the specific boundary conditions \cite{Giombi:2008vd}.

Here, we focus on the contribution arising from the Schwarzian sector, which corresponds to the physical transverse and traceless excitations identified in the previous section. These modes are the only ones that go to zero as $T\to0$; hence, we neglect continuous modes. In Appendix~\ref{sec:vector}, we analyze the vector field case analogously to the spin-2 field, showing in particular that the ghost fields do not contribute to the Schwarzian sector. Thus, the Schwarzian contribution is determined solely by the functional determinant of the Lichnerowicz operator
\begin{equation}
    Z_{\text{Sch}} = \int \mathcal{D}h_{\mu\nu} \,e^{\frac{1}{64 \pi G_N} \int dx^3\, \sqrt{g} \,h^{\mu \nu} (\nabla^2+2) h_{\mu \nu}} = \left[ \det\left(\nabla^2+2\right) \right]^{-1} 
\end{equation}
Since we have explicitly constructed the spectrum of normalizable modes for this operator in \eqref{schwarzian_spectrum}, and treating the modes $n$ and $-n$ as independent variables in the path integral measure\footnote{In the analysis of near-horizon near-extremal black holes, the modes $n$ and $-n$ are usually not independent, as they satisfy $h_{\mu\nu}^{(-n)}=(h_{\mu\nu}^{(n)})^*$ \cite{Toldo:2024,Joacokerr}. In the present case, the modes $h_{\mu\nu}(k_{\text{sch}})$ obtained in \eqref{graviton_schwarzian_z} do not satisfy this condition in the full geometry approach; however, they do satisfy it when taking the NHNE limit (see \eqref{schwarian_NHNE}).}, the functional determinant reduces to a product over the eigenvalues $\alpha_n$. The modes contributing to the partition function are restricted to the normalizable sector that satisfies the boundary conditions given by \eqref{bcc}. Hence, the Schwarzian contribution reads
\begin{align}\label{Z_sch}
    Z_{\text{Sch}}=\prod_{|n|=2}^{[\frac{2\,r_+}{r_+-r_-}]} \Big(\frac{\alpha_n}{64 \pi G_N}\Big)^{-1}
\end{align}
where $\alpha_n$ is the spectrum obtained for the modes $k_{\text{sch}}$ in \eqref{schwarzian_spectrum_2}, and the product runs over the range determined by the normalizability condition and the boundary conditions.

Next, we calculate the free energy in the grand canonical ensemble,
\begin{equation}
    \Phi_{Sch}= T \sum_{|n|=2}^{[\frac{2\,r_+}{r_+-r_-}]} \log\left(\frac{\alpha_n}{64 \pi G_N}\right)
\end{equation}
In the grand canonical ensemble, we fix the ratio $\Omega=r_-/r_+$ (so that $r_-=\Omega\,r_+$). Expressing the spectrum $\alpha_n$ in terms of $\Omega$, the grand potential becomes
\begin{equation}
    \Phi_{Sch}=T\sum_{|n|=2}^{[\frac{2}{1-\Omega}]} \log \left(\frac{|n|(1-\Omega)\left(2-|n|(1-\Omega)\right)}{64 \pi G_N}\right)
\end{equation}
and the entropy is given by
\begin{equation}
S_{Sch}= -\partial_T\Phi_{Sch}=- \sum_{|n|=2}^{\frac{2}{1-\Omega}} \log \left(\frac{|n|(1-\Omega)\left(2-|n|(1-\Omega)\right)}{64 \pi G_N}\right)
\end{equation}
Recall that the only modes satisfying the boundary conditions and the ultralocal measure are the Schwarzian modes (thoroughly analyzed in Sec.~\ref{sec:Schwarzian}) and the non-zero modes (obtained in Sec.~\ref{sec:normalizable_modes}). Regarding the latter, their asymptotic behavior follows \eqref{bcc} with $m=1+r_-/r_+$, thus satisfying the boundary conditions. Therefore, we conclude that the full $I_{TT}$ term is exhausted by these two sectors, i.e., $Z_{TT}=Z_{\text{Sch}}\,Z_{\text{nz}}$, where the non-zero mode contribution to the free energy can be straightforwardly computed using the spectrum $\alpha_n^{\text{(nz)}}=1+k_n^{\text{(nz)}\,2}$ given in \eqref{non_zero_spectrum}. However, since $\alpha_n^{\text{(nz)}}\neq0$ when $r_-\to r_+$, the Schwarzian contribution dominates in the near-extremal limit.

Finally, in order to match with known results in the near-horizon near-extremal (NHNE) limit, we approach the throat by expanding $r_\pm=r_0\pm \pi\, T/2$, so that $\Omega=1-\pi\,T/r_0$. Taking this limit in the Schwarzian sector of the partition function \eqref{Z_sch} and computing the entropy analogously, we finally obtain the Schwarzian correction to the entropy
\begin{equation}
S^{\text{NHNE}}_{Sch}=   \log \prod_{|n|=2}^\infty \frac{32 r_0 \pi G_N}{ |n| T}=\frac32\log  \left(\frac{T}{32 r_0 \pi G_N (2 \pi)^{\frac13} r_0}\right)
\end{equation}
where we have employed zeta function regularization as in \cite{Kapec:2024zdj}.

\newpage

\section{Schwarzian modes from the Kerr-Schild metric}\label{sec:Kerr_Schild}

In this section, we show that the Schwarzian modes found in Sec.~\ref{sec:Schwarzian} can also be obtained from the Kerr–Schild form of the Euclidean BTZ metric. The goal of this derivation from a purely geometric perspective is to develop tools for identifying Schwarzian modes in the full geometry of higher-dimensional black holes, since direct perturbative methods fail to yield analytical solutions \cite{JOACO}.

The Kerr–Schild metric provides a systematic way to generate new gravitational solutions from a simple base metric \cite{Carrillo:2019,Taub:1981}. The underlying idea is to describe perturbations that propagate along null directions already defined by the background geometry, preserving its causal structure (for a complete review, see \cite{Stephani:2003}). Here we define the Kerr-Schild metric in curved base space\footnote{We keep the notation $g_{\mu\nu}$ for the Euclidean BTZ metric, i.e. the base metric.} \cite{Taub:1981}
\begin{equation}\label{Kerr_Schild_metric}
    \tilde{g}^{\mu\nu}= g^{\mu\nu} + \chi\: k^\mu k^\nu
\end{equation}
where $\chi$ is a scalar field and $k^\mu$ is a null and geodesic vector with respect to both the BTZ and the Kerr-Schild metrics
\begin{equation}
    \tilde{g}_{\mu\nu} k^\mu k^\nu=g_{\mu\nu}k^\mu k^\nu=0 \qquad\&\qquad \tilde{\nabla}_\mu k^\mu=\nabla_\mu k^\mu=0 
\end{equation}
where $\tilde{\nabla}_\mu$ is the covariant derivative defined by $\tilde{g}_{\mu\nu}$. 

Defining $k_\mu\equiv g_{\mu\nu}k^\nu$, the inverse metric takes the form
\begin{equation}
    \tilde{g}_{\mu\nu}= g_{\mu\nu} - \chi\: k_\mu k_\nu
\end{equation}

We now search for a null, geodesic vector with respect to the Euclidean BTZ metric \eqref{BTZ_metric}. This metric has two Killing vectors~\cite{BTZ}, $\xi_t =\partial_t$ and $\xi_\phi =\partial_\phi$, leading to two conserved quantities along the geodesic
\begin{equation}\label{constants}
    \begin{aligned}
        E&= g_{\mu\nu}\xi_t^\mu \dot{x}^\nu =\left(f(r)-\frac{r_+^2 r_-^2}{r^2}\right)\dot{t} +i\:r_-r_+\dot{\phi}
        \\
        L&= g_{\mu\nu}\xi_\phi^\mu \dot{x}^\nu = i\:r_+r_-\dot{t} +r^2\dot{\phi}
    \end{aligned}    
\end{equation}
where the dot denotes differentiation with respect to an affine parameter $\lambda$ ($\dot{x}\equiv\frac{dx}{d\lambda}$). Then, for a null geodesic we have:
\begin{equation}
    g_{\mu\nu} \dot{x}^\mu\dot{x}^\nu = f(r) \:\dot{t}^2+\frac{\dot{r}^2}{f(r)}+r^2\left(\dot{\phi}+i \frac{r_+ r_-}{r^2}\:\dot{t}\right)^2 =0
\end{equation}
Using the conserved quantities \eqref{constants}, one finds
\begin{equation}
    \dot{r}=\pm i\sqrt{\left(E-i\frac{r_- r_+}{r^2}L\right)^2 +\frac{L^2}{r^2}f(r)}
\end{equation}
The general solution of these geodesics has been studied in detail in \cite{Farina:1993,Cruz:1994}; here we are only interested in the corresponding null vector $k^\mu$. In this Euclidean version of the geodesic, the sign $\pm$ does not have a clear physical interpretation. Upon Wick-rotating back to Lorentzian signature via $t_E\to i\:t_L$, the sign $+$ corresponds to an outgoing geodesic toward the horizon, while the sign $-$ describes an ingoing one (this depend on the chosen Wick rotation). In what follows, we will consider only the minus sign. Then, taking $k^\mu\equiv \dot{x}^\mu$, we obtain
\begin{equation}
    k_\mu dx^\mu = E\:dt -\frac{i}{r^2 f(r)}\sqrt{L^2 r^2 f(r)+\left(E\: r^2-i \: L\: r_- r_+ \right)^2} \:dr +L\:d\phi 
\end{equation}
It turns out that this null vector, which is null with respect to the base metric, it is also null under the Kerr–Schild metric. Moreover, it satisfies the geodesic equations for both $g_{\mu\nu}$ and $\tilde{g}_{\mu\nu}$ simultaneously. Thus, $k^\mu$ is indeed the general null and geodesic vector for a BTZ base metric.

Having identified the underlying null and geodesic vector in the Kerr-Schild construction, note that it naturally defines a congruence of null geodesics (i.e. a null hypersurface) through its integral curves. 
In other words, the congruence is entirely determined by the flow generated by $k^\mu$, which provides the tangent direction to the geodesics at each spacetime point. The evolution of such a congruence is governed by the Raychaudhuri equation (see \cite{Blau:2025} for a detail review on null geodesic congruences). The importance of the Raychaudhuri equation is that it dictates the focusing of horizon generators and its relation to changes in horizon area due to matter fluxes. This provides the geometric foundation of the singularity and area theorems~\cite{Hawking_Ellis:1973,Wald:1984}.

For a null geodesic in $d$ dimensions, the transverse space has dimension $d-2$. Then, in $d=3$ there can no be shear neither twist,  since they cannot be supported in the transverse space. The Raychaudhuri equation therefore simplifies to
\begin{equation}
    \frac{d\theta}{d\lambda}=-\theta^2 -R_{\mu\nu}k^\mu k^\nu
\end{equation}
where $\theta(\lambda)\equiv\nabla^\mu k_\mu$ and $\lambda$ the affine parameter. Recall that $\theta$ is the expansion of the generators, which vanishes identically for null hypersurfaces such as horizons, then the equation could be solve for $\theta(\lambda)$. However, unlike the Killing horizon which is null only in the horizon hypersurface, the vector $k^\mu$ is globally null. Moreover, since the underlying base geometry is maximally symmetric, $R_{\mu\nu}\propto g_{\mu\nu}$, the Ricci contraction $R_{\mu\nu}k^\mu k^\nu$ vanishes identically. 
Lastly, because the geodesic is tangent to a null hypersurface at every point, there is no expansion of the generators, $\frac{d\theta}{d\lambda}=0$, implying\footnote{In principle, one could also allow a non-trivial solution of the form
\begin{equation*}
    \frac{d\theta}{d\lambda} =-\theta^2 \quad \Longrightarrow \quad \theta(\lambda)=\frac{1}{\lambda-\lambda_0}
\end{equation*}
which diverges at finite affine time. This means that all the null geodesics focus at finite time, leading to a singular congruence. Hence, the only smooth congruence defining a regular null hypersurface at all times is obtained for $\theta=0$. 
However, the argument used above is more straightforward to generalize to higher-dimensional black holes.} 
\begin{equation}
    \theta=\nabla_\mu k^\mu=0 \quad\Longrightarrow\quad L=\pm i\:E
\end{equation}
The resulting vector is then 
\begin{equation}
    k_\mu^{(\pm)} dx^\mu  = -i\:|E|\left(\frac{(r_+ \pm r_-)}{r f(r)}\:dr +\frac{E}{|E|}(i\:dt \mp d\phi) \right)
\end{equation} 
where the upper and lower signs correspond to choosing $L=\pm i\:E$, respectively. Thus, there are two possible null and geodesic vectors $k_\mu^{(\pm)}$ for the Kerr-Schild construction. Notice that $k_\mu^{(+)}$ coincides exactly with the linear factorization $K_\mu$ of the Schwarzian mode found in the analysis of BTZ quantum perturbations in \eqref{graviton_factorize}, while $k_\mu^{(-)}$ matches the non-zero mode \eqref{graviton_nonschwarzian}. In this case, the physical interpretation of the sign can be understand by considering the extremal limit $r_-\to r_+$\footnote{In Lorentzian signature, the Killing horizon generator is $\chi_L=\partial_{t_L} +\frac{r_-}{r_+}\partial_\phi$. Under Wick rotation $t_L\to-i\: t_E$, this becomes in Euclidean signature $\chi_E=i\:\partial_{t_E} +\frac{r_-}{r_+}\partial_\phi$. Then, the Euclidean geodesic defining the horizon satisfies $0=i\:E +\frac{r_-}{r_+} L$, which in the extremal case implies $L=-i\:E$}: $k_\mu^{(-)}$ coincides with the Killing horizon and fixes the hypersurface at the horizon, while $k_\mu^{(+)}$ corresponds to another null vector that diverges at the horizon and carries opposite angular momentum to that of the horizon hypersurface. 

Having determined $k_\mu^{(\pm)}$, it remains to obtain the scalar field $\chi$ in the Kerr-Schild metric \eqref{Kerr_Schild_metric}. This can be obtained by taking the transverse gauge condition,
\begin{equation}\label{k_transverse}
    \nabla^\mu \left(\chi k_\mu^{(\pm)} k_\nu^{(\pm)}\right)=0 \quad\Longrightarrow\quad k_\mu^{(\pm)} \nabla^\mu \chi=0 
\end{equation}
where we have used that, as argued above, $k_\mu$ is divergence-free (hence $\nabla^\mu k_\mu=0$). Consequently, the equation for $\chi$ reduces to a simple first-order differential equation. Using the ansatz $\chi(t,r)=e^{i E\,t}\,R(r)$, one immediately finds
\begin{equation}\label{KS_scalar}
    \chi^{\pm}= \mathcal{N}_0\: e^{i E\: t} \left(\left(r^2-r_- ^2\right)^{-r_- } \left(r^2-r_+ ^2\right)^{r_+ }\right)^{\frac12\frac{|E|
   }{r_+^2-r_-^2}}
\end{equation}
Requiring regularity and periodicity in Euclidean time restricts the energy $E$ to the discrete Matsubara frequencies, $E=2\pi\,n\,T$. Upon this identification, the scalar field \eqref{KS_scalar} coincides (up to a normalization constant) with the scalar profiles obtained for the Schwarzian and non-zero modes in \eqref{scalar_graviton} and \eqref{graviton_nonschwarzian}, respectively. Furthermore, substituting this Matsubara identification into the null geodesic vectors $k^{(\pm)}_{\mu}$, we find that the full combination $\chi^\pm\,k_\mu^{(\pm)} k^{(\pm)}_\nu$ exactly reproduces the one-loop gravitational perturbations studied in Sec.~\ref{sec:normalizable_modes}. In particular, the Schwarzian mode of Sec.~\ref{sec:Schwarzian} emerges here from a purely geometric construction.

The correspondence between perturbative gravity and gauge theories is known as the \textit{double copy} \cite{Monteiro:2014}. This framework allows one to interpret the BTZ black hole as a gauge field configuration in an $AdS_3$ background \cite{Carrillo:2018}. In our case, however, the double copy involves an Abelian gauge field defined on the BTZ geometry, where the associated gauge field reproduces the Schwarzian modes of the black hole. Moreover, the classical copy $A_\mu\equiv\chi\,k_\mu$ yields a Schwarzian mode of the vector field (see Appendix~\ref{sec:vector}). Therefore, there exists a close relationship between the Schwarzian modes on the gauge theory and gravity sides, which can be understood in terms of the double copy construction.

Beyond the specific case of three-dimensional gravity, the observations made in this section highlight the power of the Kerr-Schild ansatz as a generator of physical Schwarzian modes. By mapping the complex second-order Lichnerowicz eigenvalue problem into a set of first-order geometric constraints on null vectors, this approach provides a robust framework for identifying Schwarzian modes. This geometric perspective may provide a useful framework for extending these results to higher-dimensional black holes—such as Kerr black holes—where the Schwarzian modes can be obtained analytically only in the near-extremal limit.

\section{Discussion and future directions}

We have presented a detailed construction of the Schwarzian modes for the BTZ black hole within the full geometry, explicitly deriving the conditions required to extract them from the general solution—details that were not fully addressed in their initial introduction in \cite{JOACO}. We also obtained the rotational modes, which are normalizable and become zero modes in the extremal limit (exhibiting the same behavior as the Schwarzian modes); however, they do not obey the required boundary conditions \eqref{bcc}. This confirms the discussion in \cite{JOACO}, where rotational modes were identified in the near-horizon analysis but were expected to be ill-defined from the perspective of the full geometry. This framework opens the possibility of extending the analysis to Schwarzian modes of arbitrary spin. The modes identified here arise from spin-2 perturbations, representing the second bosonic spin contributing to the Schwarzian sector, as we have shown that vector perturbations indeed support such modes. Moreover, it would be interesting to apply the same method to higher-spin bosonic fields and their fermionic counterparts; to our knowledge, it is currently being explored the latter direction in~\cite{Castro:2026}, this will clarify the supergravity case.

Throughout the paper, we showed that the existence of Schwarzian modes is intrinsically connected to the fact that these modes are also exact solutions of the corresponding linearized equations of motion — a point that, to our knowledge, had not been emphasized before in the literature. In particular we showed that there is a double copy relationship between the Schwarzian vector fields and the gravitational Schwarzians modes.

We found that the Schwarzian modes from one-loop gravitational perturbations can also be derived from a purely geometrical perspective using the Kerr–Schild metric. In three-dimensional gravity, the equivalence between perturbative and exact Kerr–Schild descriptions is natural, as the theory is one-loop exact. Nevertheless, our arguments rely solely on geometrical considerations, suggesting that a similar analysis could be performed for higher-dimensional spacetimes — such as Kerr or Reissner–Nordström black holes — where Schwarzian-like modes are known to appear, sometimes only numerically, in the near-horizon region.

As a final remark, we showed a connection between the Schwarzian modes and the double copy framework, this is different from the usual double copy paradigm \cite{Beetar:2024ptv} as we are relating quantum corrections on the vector and gravitational side. We have shown that the Schwarzian modes naturally emerge from a Kerr–Schild ansatz with an Abelian Yang–Mills field. It will be interesting to understand if this relationship also hold for higher dimensional black holes , and there might be a possibly in relation with the double field theory framework \cite{Lescano:2024gma}.  

\bigskip

\section*{Acknowledgements}

We would like to thank {\it Siembra-HoLAGrav} and the {\it GravUC School 2024: Topics on Quantum Gravity} for providing a stimulating environment that inspired the beginning of this work. We are especially grateful to Alejandra Castro, Adam Bac and Cristóbal Corral for their valuable insights. It is also a great pleasure to acknowledge discussions with Leopoldo Pando Zayas, Chiara Toldo, Joaquin Turiaci, Nicolas Grandi, Guillermo Silva, Diego Correa and Juan Maldacena. 

Lucas Acito and Matías Nicolás Sempé are supported by a CONICET fellowship. 

\newpage

\appendix

\section{Schwarzian modes in vector field}\label{sec:vector}

In this Appendix, we show that a general transverse vector field possesses Schwarzian-like modes. However, for the specific case of the ghost field, these normalizable modes do not behave as zero modes in the extremal limit. Consequently, the ghost field contribution to the one-loop path integral computation in Sec.~\ref{sec:1loop} is not dominant in the near-extremal limit. First, we compute the normalizable vector modes following the same analysis detailed in Sec.~\ref{sec:spin2}. In three dimensions, the relevant differential operator acting on the vector field is given by \cite{Leston, Indios}
\begin{equation}    
    (-\Box+2) A_\mu=(k^2+4) A_\mu
    \label{so}
\end{equation}
As in the spin-2 case, this second-order equation can be factorized into a first-order operator acting twice on the field \cite{Vasiliev:1997}, namely
\begin{equation}
    \epsilon_{\mu}^{\:\nu \rho} \nabla_\nu A_\rho=\pm k\: A_\mu 
    \label{fo}
\end{equation}
which also implies the gauge fixing condition
\begin{equation}
    \nabla^\mu A_\mu=0
\end{equation}
From the $z$-component of \eqref{fo}, we immediately obtain
\begin{equation}\label{A_z}
    A_z= \pm\frac{\partial_- A_+ -\partial_+ A_-}{2\:k \:z}
\end{equation}
Hence, only two independent components remain, namely $A_+$ and $A_-$, from which $A_z$ is determined. Expanding the Laplacian in these components gives
\begin{equation}
    \begin{pmatrix}
        \Box A_{+}\\
        \Box A_{-}
    \end{pmatrix} = \begin{pmatrix}
        \Delta A_{+}\\
        \Delta A_{-}
    \end{pmatrix} + \begin{pmatrix}
        -2 & \mp2\:k  \\
        \pm2\:k & -2 
    \end{pmatrix} \begin{pmatrix}
        A_{+}\\
        A_{-}
    \end{pmatrix} 
\end{equation}
where the upper/lower signs correspond to the $\pm$ choice in \eqref{fo}, and the scalar Laplacian is defined as $\Delta A_{\mu} =\partial_\alpha (\sqrt{g}\,\partial^\alpha A_{\mu})/\sqrt{g}$. This system can be diagonalized by introducing the combinations
\begin{equation}\label{A_trans}
    \left\{\begin{aligned}
        A_-&=i\left(A_2-A_1\right)
        \\
        A_+&=A_1+A_2
    \end{aligned}\right.
\end{equation}
which leads to the decoupled equations
\begin{equation}
    \left\{\begin{aligned}
        \Delta  A_1 +k\left(k-2\:i\right)A_1 =&0
        \\
        \Delta  A_2 +k\left(k+2\:i\right) A_2 =&0
    \end{aligned}\right.
\end{equation}
Thus, the problem reduces to two scalar field equations. These can be solved using the same ansatz as in the spin-2 case,
\begin{equation}
    A_j=e^{i (\kappa_+ x_++\kappa_- x_-)} R_j(z)
    \label{ansatzvec}
\end{equation} 
with $\kappa_-=n$ and $\kappa_+=-i\, n\,r_-/r_+$, in order to have Matsubara modes with $n\in\mathbb{Z}$ as in Sec.~\ref{sec:normalizable_modes}. Taking the regular solution at $z\to0$, we find
\begin{equation}\label{vector_scalars}
    \begin{aligned}
        R_1&=e_1\: z^{\frac{|n|}2} (1-z)^{-\frac{ik}{2}} \, _2F_1\left[a-\frac{ik}{2},b-\frac{ik}{2},c,z\right]
        \\
        R_2&=e_2\: z^{\frac{|n|}2} (1-z)^{\frac{ik}{2}} \, _2F_1\left[a+\frac{ik}{2},b+\frac{ik}{2},c,z\right]
    \end{aligned}
\end{equation}
where $e_1$ and $e_2$ are polarization constants, and we defined the parameters
\begin{equation}\label{vector_param}
    a=\frac{|n|}{2}\left(1-\frac{r_-}{r_+}\right)\:,\qquad b= \frac{|n|}{2}\left(1+\frac{r_-}{r_+}\right)\:,\qquad c=|n|+1
\end{equation}
Substituting this solution into \eqref{fo}, we obtain for the polarizations
\begin{equation}\label{vector_pol}
    \left(i\:k+|n|\left(-1\pm\frac{r_-}{r_+}\right)\right)e_1 =\left(i\:k+|n|\left(1\pm\frac{r_-}{r_+}\right)\right)e_2
\end{equation}
where the top/bottom signs correspond to the $\pm$ choice in \eqref{fo}, respectively. Notice that, as in the spin-2 case, the radial part $R_i(z)$ of the scalar fields depends only on $|n|$, including the polarization constants.

We have found the solutions that satisfy both the quadratic \eqref{so} and linear operator \eqref{fo}. Following the discussion for the spin-2 field in Sec.~\ref{sec:normalizable_modes}, we now identify the normalizable modes by computing the norm
\begin{equation}\label{vector_norm}
    ||A||^2= \int d^3x \:\sqrt{g} \:A_{\mu}^{(n)}A^{\mu\,(-n)} = \int d^3x \:\frac2z\left(R_1^2 \left(\frac{b^2}{k^2} -\frac14 \right)+\frac{R_1\:R_2}{1-z}  +R_2^2 \left(\frac{a^2}{k^2} -\frac14 \right)\right)
\end{equation}
where we have used \eqref{A_z} and \eqref{A_trans} to write it in terms of the radial part $R_i(z)$ of the scalar fields \eqref{ansatzvec} (which only depends on $|n|$), and the parameters $a$ and $b$ were defined in \eqref{vector_param}. Hence, we write the norm as a combination of smooth functions, so it suffices to analyze their asymptotic behavior to determine convergence.

We next examine the near-horizon ($z\to0$) and near-boundary ($z\to1$) behaviors. In the near-horizon limit, the modes are regular and approach zero for $|n|>0$ without requiring any additional conditions. On the other hand, near the boundary $(z\to1)$, using the hypergeometric expansion \eqref{asymp}, the radial functions behave as
\begin{equation}\label{vector_scalars_asym}
    \begin{aligned}
        R_{1}\sim\:& e_1\: (1-z)^{-\frac{ik}{2}} \left(\frac{\Gamma(1+ik)\:\Gamma(|n|+1)}{\Gamma(c-a+\frac{ik}2)\:\Gamma(c-b+\frac{ik}2)} +\frac{\Gamma(-1-ik)\:\Gamma(|n|+1)}{\Gamma(a-\frac{ik}2)\:\Gamma(b-\frac{ik}2)}(1-z)^{1+ik}\right)
        \\
        R_{2}\sim\:& e_2\: (1-z)^{\frac{ik}{2}} \left(\frac{\Gamma(1-ik)\:\Gamma(|n|+1)}{\Gamma(c-a-\frac{ik}2)\:\Gamma(c-b-\frac{ik}2)} +\frac{\Gamma(-1+ik)\:\Gamma(|n|+1)}{\Gamma(a+\frac{ik}2)\:\Gamma(b+\frac{ik}2)}(1-z)^{1-ik}\right)
    \end{aligned}
\end{equation}
Inserting these asymptotic behaviors into the norm density \eqref{vector_norm}, normalizable modes require that the leading term decays faster than $\mathcal{O}((1-z)^{-1})$ near the boundary. In particular, as in the spin-2 case, a logarithmic divergence arises from
\begin{equation}\label{vector_log_divergence}
    \begin{aligned}
        \frac{R_{1}\:R_{2}}{1-z}\sim e_{1}\:e_{2}\:\frac{\Gamma(1+ik)\:\Gamma(|n|+1)}{\Gamma(c-a+\frac{ik}2)\:\Gamma(c-b+\frac{ik}2)}\frac{\Gamma(1-ik)\:\Gamma(|n|+1)}{\Gamma(c-a-\frac{ik}2)\:\Gamma(c-b-\frac{ik}2)}\:(1-z)^{-1}
    \end{aligned}
\end{equation}
This divergent term is independent of $k$, and it is the only term of this form in \eqref{vector_norm}. In principle, there are two ways to cancel it: (\textit{i}) set either of the polarizations $e_{1}$ or $e_{2}$ to zero; or (\textit{ii}) impose that the corresponding coefficient involving Gamma functions vanishes. However, since $c-a>1$ and $c-b>1$, the second option do not work, as the remaining exponents in the norm would be divergent. Therefore, as in the spin-2 case, only the first option leads to genuinely normalizable modes. Moreover, setting the polarizations to zero requires purely imaginary modes; thus, only these modes can lead to normalizability (i.e., satisfy the ultralocal measure \eqref{vector_norm}; the $k\in\mathbb{R}$ case was also discussed in~\cite{Leston}, but as explained in Sec.~\ref{sec:normalizable_modes}, this does not contradict our result).

\begin{figure}
	\begin{center}
	\includegraphics[width=0.5\textwidth]{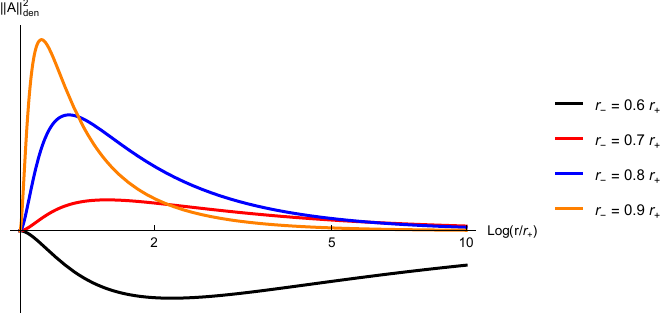}%
	\includegraphics[width=0.5\textwidth]{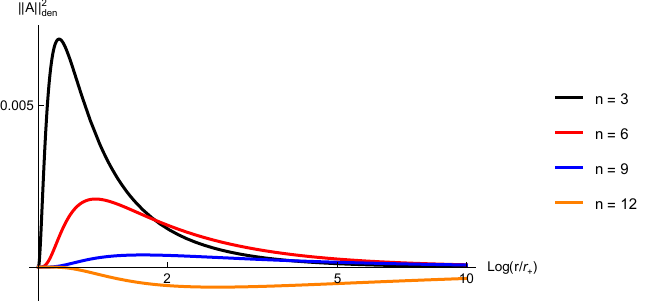}
    \caption{Norm density of the vector modes, $||A||^2_{\sf den}=\sqrt{g}\,A_{\mu}^{(n)}A^{\mu\,(-n)}$, as a function of the proper radial coordinate $\log(r/r_+)$. \textbf{Left}: Fixed mode $n=3$ for different ratios $r_-/r_+$. The modes become increasingly localized near the horizon as extremality is approached ($r_- \to r_+$), but they do not become zero modes. Away from extremality, the modes spread out and eventually become non-normalizable, as predicted by \eqref{vector_norm2}, which manifests as a negative norm density. \textbf{Right}: Norm density for several modes at fixed $r_-=0.9\,r_+$. In agreement with \eqref{vector_norm2}, only a finite number of modes are normalizable; in this case, modes with $n \geq 11$ become non-normalizable, while the lowest modes are increasingly localized near the horizon.}
    \label{fig:vector}
	\end{center}
\end{figure}

\paragraph{Normalizable modes.} As mentioned above, normalizability requires the leading term in \eqref{vector_norm} to decay faster than $(1-z)^{-1}$. In particular, setting the polarizations $e_{1}$ or $e_{2}$ to zero using \eqref{vector_pol} fixes the dependence of $k$ on the temperature
\begin{equation}\label{vector_modes}
    k_{1}=-i\:|n|\left(1-\frac{r_-}{r_+}\right) \qquad \text{or}\qquad k_{2}=i\:|n|\left(1+\frac{r_-}{r_+}\right)
\end{equation}
where $k_{1}$ arises from setting $e_1=0$, while $k_{2}$ corresponds to $e_2=0$ (here we fixed the plus sign in \eqref{fo} to simplify notation). Since $|k_2|>2$, it follows from the asymptotic behavior of $R_1$ in \eqref{vector_scalars_asym} that this mode is non-normalizable. Thus, the only normalizable mode for the vector field is associated with $k_1$, for which we obtain from \eqref{vector_scalars}
\begin{equation}\label{vector_mode1_z}
    A_{\mu}^{(n)}\,dx^\mu=\frac{1}{2}\:\mathcal{N}_{1}\:e^{i\:2\pi nT\:t}\:z^{\frac{|n|}{2}-1} (1-z)^{\frac{|n|}{2}(\frac{r_-}{r_+}-1)} \left(dz+2 z\,\frac{n}{|n|}(i\,dt-d\phi) (r_+-r_-)\right)
\end{equation}
Integrating these modes, we obtain the norm \eqref{vector_norm}
\begin{equation}\label{vector_norm2}
    ||A||^2 = \frac{4\pi^2\,(\mathcal{N}_{1})^2\:r_+\,\Gamma(|n|) \,\Gamma\left(|n|\left(\frac{r_-}{r_+}-1\right)+1\right)}{\Gamma \left(|n|\frac{r_-}{r_+}+1\right)}\,,\qquad 0<|n|<\frac{r_+}{r_+-r_-}
\end{equation}
Thus, we can define the normalization as 
\begin{equation}\label{vector_norm1_cte}
    (\mathcal{N}_{1})^{2}= \frac{\Gamma \left(|n|\frac{r_-}{r_+}+1\right)}{4\pi^2\:r_+\,\Gamma(|n|) \,\Gamma\left(|n|\left(\frac{r_-}{r_+}-1\right)+1\right)}
\end{equation}
The vector field solution \eqref{vector_mode1_z} exhibits characteristics similar to those of the Schwarzian modes discussed in Sec.~\ref{sec:Schwarzian}: the number of normalizable modes is finite but diverges in the extremal limit, which can be interpreted as localization near the black hole throat (see Fig.~\ref{fig:vector}). However, for the spectrum of the quadratic operator \eqref{so} required to compute the 1-loop path integral in Sec.~\ref{sec:1loop}, we obtain
\begin{equation}    
    (-\Box+2) A_\mu^{(n)}=\beta_n\, A^{(n)}_\mu
\end{equation}
with
\begin{equation}
    \beta_n=4-n^2\left(1-\frac{r_-}{r_+}\right)^2\,,\qquad 0<|n|<\frac{r_+}{r_+-r_-}
\end{equation}
Notice that the eigenvalue $\beta_n$ never vanishes for the normalizable modes found. We thus conclude that these transverse vector fields (identified as ghost fields) admit normalizable modes, but they are not Schwarzian modes for the functional determinant that appears in \eqref{so} and, consequently, do not contribute to the logarithmic $\log\,T$ correction in the three-dimensional gravitational partition function. Nevertheless, there exist Schwarzian modes whose eigenvalues vanish at extremality in the case of the massless vector field. This implies a double copy relationship between the Schwarzian modes of the graviton and those of the massless vector field.

\newpage

\bibliographystyle{JHEP}
\bibliography{references}

\end{document}